\documentclass[11 pt]{article}
\usepackage[utf8]{inputenc}
\usepackage[a4paper, total={7in, 10in}]{geometry}
\usepackage{graphicx}
\usepackage{booktabs} 
\usepackage{tabularx} 
\usepackage{amsmath}
\usepackage{amssymb}
\usepackage{multirow}

\usepackage{authblk}
\title{On the Cooling of Compact Stars in Light of the HESS J1731-347 Remnant}
\author{Dimitrios G. Nanopoulos\footnote{dnanopou@physics.auth.gr}}
\author{Pavlos Laskos-Patkos\footnote{plaskos@physics.auth.gr}}
\author{Charalampos C. Moustakidis\footnote{moustaki@auth.gr}}
\affil{Department of Theoretical Physics, Aristotle University of Thessaloniki, 54124 Thessaloniki, Greece}

\begin{document}

\maketitle

\abstract{Recent analyses on the central compact object in the HESS J1731-347 supernova remnant reported not only surprising structural properties (mass $M$ and radius $R$), but also an interesting thermal evolution. More precisely, it has been estimated that $M=0.77^{+0.20}_{-0.17}M_\odot$ and $R=10.4^{+0.86}_{-0.78}$ km (at the $1\sigma$ level), while a redshited surface temperature of $153^{+4}_
{-2}$ keV at an age of 2-6 kyrs has been reported. In the present work, we conduct an in-depth investigation on the possible nature (hadronic, hybrid, quark) of this compact object by attempting to not only explain its mass and radius but also the corresponding estimations for its temperature and age. In the case of hybrid stars we also examine possible effects of the symmetry energy on the activation of different neutrino emitting process, and hence on the resulting cooling curves. We found that the reported temperature and age may be compatible to hadronic stellar configurations regardless of whether pairing effects are included. In the scenario of hybrid stars, we found that the strange quark matter core has to be in a superconducting state in order to reach an agreement with the observational constraints. In addition, the hadronic phase must be soft enough so that the direct Urca process is not activated. Furthermore, we have shown that the considered cooling constraints can be reconciled within the framework of strange stars. However, quark matter has to be in a superconducting state and the quark direct Urca process needs to be blocked.}


\section{Introduction} \label{sec1}

One of the major unresolved questions in nuclear astrophysics is related to the nature of matter in the cores of compact stars. In particular, the stellar interior could be purely hadronic (nucleons and potentially hyperons), but the extreme densities may allow for the presence of exotic forms of matter such as deconfined quarks~\cite{Witten-1984, Annala-2020}. In that direction, theorists have proposed the possible existence of strange quark stars~\cite{Weber-2005}, composed solely of strange quark matter (SQM), and hybrid stars (HSs)~\cite{Heiselberg-2000}, where a SQM core is covered by a layer of hadronic matter. Interestingly, different assumptions about the nature of compact objects yield distinct predictions about their properties~\cite{Lattimer-2001}. In that sense, future and precise astronomical measurements may provide the opportunity for a glimpse in the interior of compact stars. It is worth noting that we do not strictly refer to the measurement of structural stellar properties such as masses and radii. In particular, fruitful information about the state of matter in the interior of compact objects can be derived by studying their evolution. For instance, signature phenomena associated with first-order phase transitions could be found in the rotational evolution of compact stars~\cite{Glendening-1997,Heiselberg-1998,Cheng-2002,Spyrou-2002}, while the impact from the presence of exotic degrees of freedom may appear through distinct energy loss mechanisms driving the thermal evolution of a stellar configuration~\cite{Page-2006,Yakovlev-2001}.

In past years, a lot of measurements on the masses and radii of different compact stars have appeared in the literature, allowing us to probe the nuclear equation of state (EOS)~\cite{Antoniadis-2013,Cromatie-2020,Romani-2022,Abbott-2018,Riley-2019,Miller-2019,Doroshenko-2022,Zhang-2011,Choudhury-2024,Salmi-2024,Vinciguerra-2024}. Recently, Doroshenko et al.~\cite{Doroshenko-2022}, reported their results on the bulk properties of the central compact object (CCO) in the HESS J1731-347 remnant, which were found to be particularly surprising (compared to previously inferred measurements on other compact objects, see Ref.~\cite{Zhang-2011} on neutron star masses). More precisely, the authors estimated that  $M=0.77^{+0.20}_{-0.17}M_\odot$ and $R=10.4^{+0.86}_{-0.78}$ km (at the $1\sigma$ level)~\cite{Doroshenko-2022}. Subsequently, this result triggered a wave of studies which attempted to interpret it by considering various different hypotheses about the nature of dense nuclear matter~\cite{DiClemente-2023,Horvath-2023,Oikonomou-2023,Das-2023,Rather-2023, Kourmpetis-2025,Tsaloukidis-2023,Sagun-2023,Laskos-2024,Laskos-2025,Li-2023n,Mariani-2024,Gholami-2024, Gao-2024, Pal-2025,Castillo-2025,Kubis-2025,Brodie-2023,Huang-2023,Li-2023,Kubis-2023,Char-2024,Tewari-2024,Zhang-2025,Yang-2024}. Specifically, it  has been suggested that the aforementioned CCO could be a strange quark star ~\cite{DiClemente-2023,Horvath-2023,Oikonomou-2023,Das-2023,Rather-2023, Kourmpetis-2025}, a HS exhibiting a first-order phase transition~\cite{Tsaloukidis-2023,Sagun-2023,Laskos-2024,Laskos-2025,Li-2023n,Mariani-2024,Gholami-2024, Gao-2024, Pal-2025,Castillo-2025,Kubis-2025}, a hadronic star described by a rather soft nuclear model ~\cite{Brodie-2023,Huang-2023,Li-2023,Kubis-2023,Char-2024,Tewari-2024,Zhang-2025} or a dark matter admixed compact star~\cite{Yang-2024}. For a relevant critique discussing some caveats of the aforementioned measurement we refer to the work of Ref.~\cite{AlfordHalpern-2023}.

Apart from the particular structural characteristics of the CCO in the HESS J1731-347 remnant, its thermal evolution has also been a subject of discussion. Specifically, the measured redshifted surface temperature of this object appears to be quite high with respect to its age. Note that the high temperature initially reported by H.E.S.S. Collaboration~\cite{HESS-2011},
$T_s^\infty=153^{+4}_
{-2}$ keV, was difficult to reconcile with an estimated age of 27 kyr~\cite{Horvath-2023}. Interestingly, the aforementioned estimation led to the discussion of the CCO's thermal evolution~\cite{Ofen-2015} prior to the results of Ref.~\cite{Doroshenko-2022}. Subsequent studies have now estimated that the CCO is probably younger with a respective age of 2--6 kyr~\cite{Horvath-2023,Acero-2015,Cui-2016,Maxted-2018}. Following that, the studies of Refs.~\cite{DiClemente-2023,Horvath-2023} have suggested that the CCO could be composed of color superconducting strange quark matter, without however performing the relevant calculations. This qualitative interpretation is motivated by the fact that the low mass and radius can be easily reconciled as the properties of a strange quark star, while pairing would be also required to ensure the slow cooling. A recent study~\cite{Yuan-2025} working on the scenario of the CCO being a color--flavor locked strange star has suggested that an additional heating mechanism (such as $r$-mode heating) may be essential to guarantee the reconciliation of its cooling behavior. In Ref.~\cite{Sagun-2023}, the authors have conducted calculations for the thermal evolution of the CCO as a hadronic and a hybrid star, highlighting that its explanation is possible only under the assumption that stellar matter is paired. It is worth mentioning that the authors of the latter study have utilized the initial estimation for the CCO's age in their work. Zhang et al.~\cite{Zhang-2025} considered the refined value for the age and also suggested that the explanation of the HESS J1731-347 CCO as a hadronic star would require the presence of superfluidity. 

The main motivation of the present work is threefold. Firstly, we want to conduct a detailed investigation on the possible (hadronic, quark, or hybrid) nature of the CCO in the HESS J1731-347 remnant by attempting to simultaneously explain its structural properties and thermal evolution. Secondly, we wish to highlight the role of different neutrino emission processes in the attempt of reconciling the cooling behavior of the aforementioned compact object. More precisely, we aim to examine the possible role of the nuclear symmetry energy regarding the processes that may be activated in the stellar interior. Finally, we aim to illustrate the model dependency of the resulting cooling curves on the considered parametrization of nucleon superfluidity in dense matter.

This paper is organized as follows. Section~\ref{sec2} includes the hadronic and quark models considered in the present study. Section~\ref{sec3} contains an extensive presentation of the theoretical formalism of compact star cooling. In Section~\ref{sec4}, we discuss our results and their implications. In addition, Section~\ref{sec5} contains a summary of our findings. {Finally, Appendix~\ref{app:A} illustrates the effect of the u and d quark masses on the activation of the quark direct Urca process.}

\section{Equation of State}\label{sec2}
\subsection{Hadronic Matter}

For the description of hadronic matter, we will rely on a rather simplistic but robust model. In particular, we are going to construct the hadronic EOS by employing the expansion for the energy per nucleon (as a function of the baryon density $n_b$) around the nuclear saturation density ($n_0$), given as~\cite{Lattimer-2023,Lattimer-2014,Piekarewicz-2009,Divaris-2024}

\begin{equation} \label{1}
    E_b(n_b,\alpha) = E_0 + \frac{K_0}{18n_0^2}(n_b-n_0)^2 +S(n_b)\alpha^2,
\end{equation}
where $\alpha=\frac{n_n-n_p}{n_b}$, while $n_n$ and $n_p$ stand for the number densities of neutron and protons, respectively. The parameters $E_0$ and $K_0$ denote the energy per baryon of symmetric nuclear matter and the incompressibility at $n_0$. Finally, $S(n_b)$ is the well-known symmetry energy which can also be expanded around $n_0$ leading to ~\cite{Lattimer-2023,Lattimer-2014,Piekarewicz-2009,Divaris-2024}
\begin{equation} \label{2}
    S(n_b) = J + \frac{L}{3n_0}(n_b-n_0)+\frac{K_{sym}}{18n_0^2}(n_b-n_0)^2,
\end{equation}
where $J$ is the symmetry energy at saturation density and $L$ is the so-called symmetry energy slope. {The parameter $K_{sym}$ is the curvature of symmetry energy, which is currently largely unknown.}

The derivation of the EOS relies on enforcing chemical equilibrium and charge neutrality to the system. These condition read 
\begin{equation} \label{3}
    \mu_n=\mu_p+\mu_e \hspace{0.5 cm}\mathrm{and} \hspace{0.5 cm} n_p=n_e,
\end{equation}
where $\mu_i$ denotes the chemical potential of different particle species and $n_e$ is the electron number density. Solving Equation~(\ref{3}) provides us with parameters $\alpha$ and  $n_e$ as a function of the baryon density. Then, the resulting EOS is given by simply summing the contributions of baryons and electrons to the energy density and pressure. The outermost layers of the stellar models are described by the well-known Baym--Pethick--Sutherland EOS~\cite{Baym-1971}. The transition density from the uniform core to the the solid crust is calculated via the Thermodynamical method~\cite{Kubis-2007,Kubis-2004,Moustakidis-2010,Moustakidis-2012}.

A discussion is appropriate regarding the selection of the present model. The main reason that this model was selected is that it allows to easily parametrize the nuclear EOS via coefficients that are experimentally attainable ($K_0$ and $L$). At this point, one might argue that such an approach may be oversimplified and cannot result in accurate representation of neutron star (NS) matter. This is, in principle, correct at large densities. However, the considered expansion should hold around the nuclear saturation density.  Therefore, given that the present study concerns the modeling of subsolar mass compact objects, for which the central density does not reach very high values, we estimate that the current choice will suffice for our analysis (see also Ref.~\cite{Sotani-2014} which demonstrates that the structural properties of a low mass NS are essentially dictated by the values of the nuclear saturation parameters). 

\subsection{Quark Matter}

The most widely-known model for the description of quark matter is the MIT bag model. In that framework, quarks stay within a bag due to the effect of an external pressure. The standard MIT bag model is then described by the following Lagrangian density~($\hbar=c=1$)~\cite{Lopes-2021a}
\begin{equation} \label{4}
    \mathcal{L}_0=\sum_{q=u,d,s} [\bar{\psi_q} (i\gamma_{\mu}\partial^{\mu}-m_q)\psi_q-B]\Theta,
\end{equation}
where $m_q$ denotes the mass of q quark, $B$ is the bag constant and $\Theta$ is a Heavyside step function which has a zero value out of the bag. For all of the models in this study
that involve strange quark matter, the values of quark masses can be considered equal to $m_u=2.3$ MeV, $m_d=4.8$ MeV and $m_s=95$ MeV, unless stated otherwise.

As performed in previous studies~\cite{Lopes-2021a,Lopes-2021b,Klahn-2015,Gomez-2016,Gomez-2019a,Gomes-2019b,Jaikumar-2021,Costantinou-2021,Zhao-2022,Costantinou-2023,Lyra-2023,Kumar-2022,Kumar-2023}, we consider an interaction between quarks through minimal coupling with a vector boson $V^\mu$. Then, the interaction term reads as~\cite{Lopes-2021a}
\begin{equation} \label{5}
    \mathcal{L}_{\rm vec}=\{-g_v\sum_{q} \bar{\psi_q} \gamma_\mu V^\mu\psi_q + \frac{1}{2}m_{V}^2 V_\mu V^\mu\}\Theta,
\end{equation}
where $g_v$ is the coupling constant of the interaction and $m_{V}$ stands for the mass of the mediating boson.~In this model, known as the vector MIT (vMIT) Bag model, the energy density can be analytically written  as~\cite{Jaikumar-2021}
\begin{equation} \label{6}
    \mathcal{E}_{Q} = \sum_{q} \mathcal{E}_q + \frac{1}{2} G_v (n_u+n_d+n_s)^2+B,
\end{equation}
where $G_{v}=(g_{v}/m_{ V})^2$ and $\mathcal{E}_q$ corresponds to the energy density of an ideal Fermi gas of q quarks, given as
\begin{equation}  \label{7}
\begin{split}
    \mathcal{E}_q  & =  \frac{3}{8\pi^2}  \Biggl\{ k_{Fq}(k_{Fq}^2+m_q^2)^{1/2}(2k_{Fq}^2+m_q^2) - m_q^4\ln \left [ \frac{k_{Fq}+(k_{Fq}^2+m_q)^{1/2}}{m_q}\right]  \Biggl\}.
\end{split}
\end{equation}
The quantity $k_{Fq}=(\pi^2n_q)^{1/3}$ stands for the q quark Fermi momentum. It is worth commenting that the formula of Equation~(\ref{6}) could be equivalently derived by the consideration of a Yukawa type potential (non Newtonian gravity model)~\cite{Yang-2021,Yang-2023}. 

Within the present model, the chemical potential of q quarks will be given by~\cite{Jaikumar-2021}
\begin{equation} \label{8}
    \mu_q=(k_{Fq}^2+m_q^2)^{1/2}+G_v(n_u+n_d+n_s),
\end{equation}
and the total pressure of the system is found through the standard thermodynamic relation
\begin{equation} \label{9}
    P_Q = \sum_{q} \mu_q n_q - \mathcal{E}_Q.
\end{equation}
Finally, the derivation of the quark EOS requires the imposition of chemical equilibrium and charge neutrality to the system. Taking into account the relevant weak processes, considering a system composed of uds quarks and electrons, the condition for chemical equilibrium reads as
\begin{equation} \label{10}
    \mu_d=\mu_u+\mu_e, \quad \mu_d=\mu_s,
\end{equation}
while the charge neutrality is expressed via
\begin{equation} \label{11}
    \frac{2}{3}n_u-\frac{1}{3}(n_d+n_s)-n_e=0.
\end{equation}
The quantities $\mu_e$, $n_e$ and $m_e$ stand for the chemical potential, number density and mass of electrons, respectively. The resulting EOS will be, once again, the sum of the contribution from quarks and electrons.

In the present work, we also considered the case where the parameter $B$, appearing in Equation~(\ref{6}), is density dependent. Following the discussion of several related papers~\cite{Burgio-2002a,Burgio-2002b,Yazdizadeh-2013,Miyatsu-2015,Sen-2023,Pal-2023}, we employed the widely used Gaussian parametrization 
\begin{equation} \label{12}
    B(n) = B_{as}+ (B_0-B_{as}) \exp
    \left[-\beta \left(\frac{n_b}{n_0}\right)^2\right],
\end{equation}
where $B_0$ and $B_{as}$ stand for the values of $B$  at zero and infinitely large density. The baryon density in quark matter is given by $n_b=(n_u+n_d+n_s)/3$. It is important to clarify that in the scenario of a density dependent bag parameter, we ought to include an additional term in the quark chemical potentials to ensure thermodynamic consistency. Notably, the chemical potential is shifted as~\cite{Kumar-2023}
\begin{equation} \label{13}
    \mu_q=(k_{Fq}^2+m_q^2)^{1/2}+G_v(n_u+n_d+n_s)+\frac{\partial B}{\partial n_q}.
\end{equation}
Then, the total pressure of the system can be once again evaluated via Equation~(\ref{9}). 

As our main intention is the study of compact star cooling, a discussion is appropriate about the inclusion of superconducting effects in the description of strange quark matter (which are known to play a crucial role in the stellar thermal evolution). We have chosen not to use the well-known color-flavor locked EOS (within the MIT bag model framework) as it does not explicitly include interactions (apart from a correction parameter that mimics the results of perturbative QCD) among quarks and it would most likely lead to difficulty at satisfying the $2M_\odot$ maximum mass constraint (in the case of hybrid stellar configurations. see Ref.~\cite{Kourmpetis-2025}). Following the discussion in the work of Ref.~\cite{Negreiros-2012} (which studies the cooling of hybrid stars), the corrections of including superconductivity to the energy density and pressure are $\sim\Delta^2\mu_B^2$ ($\Delta$ is the superconducting gap and $\mu_B$ the baryon chemical potential). Notably,  these corrections have noticeable effects {on} the structure of compact stars only when $\Delta\gtrsim50$ MeV (see Ref.~\cite{Negreiros-2012} and references therein). Therefore, by restricting the present study to lower $\Delta$ we can safely neglect the effects of superconductivity in the resulting EOS.

\subsection{Phase Transition}
In the present study, hybrid EOSs are derived through the Maxwell construction, which is the favored approach in the scenario of a large surface tension in the hadron-quark interface ($\sigma\gtrsim40$ MeV fm$^{-2}$~\cite{Yasutake-2016,Mariani-2017}). In this case, the phase transition onset can be calculated using the following equations~\cite{Bielich-2020}
\begin{equation} \label{14}
    P^I=P^{II} \hspace{0.5 cm} \mathrm{and}  \hspace{0.5 cm} \mu_{\rm B}^I=\mu_{\rm B}^{II},
\end{equation}
where the superscripts $I$ and $II$ indicate the two different phases of matter, $P$ is the pressure and $\mu_B$ the baryon chemical potential. Notably, under the Maxwell construction, charge neutrality is imposed locally leading to a discontinuity in the energy density (energy density jump). This property gives rise to several interesting possibilities such as the existence of twin star solutions (objects of the same mass but different radius)~\cite{Gerlach-1968,Kampfer-1981a,Kampfer-1981b,Glendenning-2000,Schertler-2000} or the emergence of $g$ mode oscillations~\cite{Sotani-2011}.

\section{Compact Star Cooling}\label{sec3}

\subsection{Cooling Processes in Quark Matter}\label{subsec31}
The neutrino-emitting processes in quark matter and their corresponding emissivities were investigated by Iwamoto in 1982 \cite{Iwamoto-1982}. The strongest neutrino-emitting processes in quark matter are the quark Direct Urca (QDU) processes, which are described by the beta decay of down and strange quarks \cite{Iwamoto-1982}
\begin{equation} \label{15}
    d\rightarrow u+e^-+\bar{\nu}_e,
\end{equation}
\vspace{-1.9em}
\begin{equation} \label{16}
    u+e^-\rightarrow d+\nu_e,
\end{equation}
\begin{equation} \label{17}
    s\rightarrow u+e^-+\bar{\nu}_e,
\end{equation}
\begin{equation} \label{18}
    u+e^-\rightarrow s+\nu_e.
\end{equation}
{The emissivities related to} the reactions above were calculated to be equal to~\cite{Iwamoto-1982}
\begin{equation} \label{19}
\mathcal{E}_\nu^{QDU-d} \simeq \frac{914}{315}\frac{G_c^2\cos^2\theta_c}{\hbar^{10}c^6}\: a_c\:p_{F_d}\:p_{F_u}\:p_{F_e}\:(k_B T)^6 , 
\end{equation}
\vspace{-0.5em}
\begin{equation} \label{20}
\mathcal{E}_\nu^{QDU-s} \simeq \frac{457\pi}{840}\frac{G_c^2\sin^2\theta_c}{\hbar^{10}c^7}\: (1-\cos\theta_{34})\:\mu_s\:p_{F_u}\:p_{F_e}\:(k_B T)^6  , 
\end{equation}
where $G_c \simeq 1.435\cdot10^{-49} \; \mathrm{erg \cdot cm^{3}}$ is the weak coupling constant, $\theta_c$ is the Cabibbo angle with $\cos^2\theta_c \simeq 0.948$, $a_c$ is the strong coupling constant, $k_B$ is the Boltzmann constant, $\hbar$ is the reduced Planck constant, $\mu_s$ corresponds to the s quark chemical potential and the angle factor $\theta_{34}$ is equal to $\theta_{34} = \theta_{13} + \theta_{14}$. The Fermi momenta for quarks and electrons as well as the angle factor $\theta_{34}$ are given below \cite{Iwamoto-1982}
\begin{equation}\label{21}
p_{F_i} = \hbar \: (\pi^2\:n_i)^{1/3} \quad \mathrm{(i = u, d, s)},
\end{equation}
\vspace{-0.5em}
\begin{equation}\label{22}
p_{F_e} = \hbar \: (3\:\pi^2\:n_e)^{1/3},
\end{equation}
\vspace{-0.5em}
\begin{equation}\label{23}
\cos\theta_{14} = (p_{F_s}^2 + p_{F_e}^2 - p_{F_u}^2) / 2 p_{F_s}  p_{F_e},
\end{equation}

\begin{equation}\label{24}
\cos\theta_{13} = (p_{F_s}^2 + p_{F_u}^2 - p_{F_e}^2) / 2 p_{F_s}  p_{F_u} .
\end{equation}
The QDU processes are activated if the Fermi momenta of quarks and electrons obey the triangle inequalities, $p_{F_d} < p_{F_u} + p_{F_e}$ for the d quark beta decay and $p_{F_s} < p_{F_u} + p_{F_e}$, $p_{F_s} > \lvert p_{F_u} - p_{F_e} \rvert$ for the s quark beta decay \cite{Page-2006, Iwamoto-1982}.

In the case where the conditions above are not satisfied, the neutrino emission in quark matter is dictated by the quark Modified Urca (QMU) processes (\ref{25})--(\ref{28}) and the quark Bremsstrahlung (QB) processes (\ref{29}) \cite{Iwamoto-1982, Blaschke-2001}
\begin{equation} \label{25}
    q + d \rightarrow q + u + e^- + \bar{\nu}_e,
\end{equation}
\vspace{-1.9em}
\begin{equation} \label{26}
    q + u + e^- \rightarrow q + d + \nu_e,
\end{equation}
\begin{equation} \label{27}
    q + s \rightarrow q + u + e^- + \bar{\nu}_e,
\end{equation}
\begin{equation} \label{28}
    q + u + e^- \rightarrow q + s + \nu_e,
\end{equation}
\begin{equation} \label{29}
    q_1 + q_2 \rightarrow q_1 + q_2 + \nu_e + \bar{\nu}_e,
\end{equation}
where $q, q_1$ and $q_2$ {arbitrarily correspond to} up, down and strange quarks. These reactions involve a second quark in order to ensure energy and momentum conservation. The emission rates of these specific processes are given by \cite{Iwamoto-1982, Zhou-2007} 
\begin{equation} \label{30}
\mathcal{E}_\nu^{QMU-d} \simeq 2.83\cdot10^{19}\:a_c^2\:\frac{n_b}{n_0}\:T_9^8 \quad \mathrm{erg \cdot cm^{-3} \cdot s^{-1}},
\end{equation}
\vspace{-0.5em}
\begin{equation} \label{31}
\mathcal{E}_\nu^{QB} \simeq 2.98\cdot10^{19}\:\frac{n_b}{n_0}\:T_9^8 \quad \mathrm{erg \cdot cm^{-3} \cdot s^{-1}},
\end{equation}
where $n_b$ corresponds to the baryon number density and $T_9$ is the temperature in units of $\mathrm{10^9\;K}$. In the present work only the d quark branch of the QMU processes was taken into consideration because{,} according to Ref.~\cite{Iwamoto-1982}, the s quark branch can generally be neglected.

\subsection{Specific Heat in Quark Matter}\label{subsec32}
The total heat capacity in quark matter is equal to the sum of the specific heat contributions {from} each particle species (u, d, s quarks and electrons). The quark and electron specific heat expressions used in the present work are \cite{Iwamoto-1982, Blaschke-2000} 
\begin{equation} \label{32}
c_{v_q} \simeq 2.5\cdot10^{20}\:\left( \frac{n_b}{n_0} \right)^{2/3}\:T_9 \quad \mathrm{erg \cdot cm^{-3} \cdot K^{-1}},
\end{equation}
\begin{equation}\label{33}
c_{v_e} \simeq 0.6\cdot10^{20}\:\left(x_e\: \frac{n_b}{n_0} \right)^{2/3}\:T_9 \quad \mathrm{erg \cdot cm^{-3} \cdot K^{-1}},
\end{equation}
where $n_0 = 0.155\cdot \mathrm{10^{39}\;cm^{-3}}$ is the nuclear saturation density and  $x_e$ is the electron fraction.

\subsection{Pairing in Quark Matter}\label{subsec33}
Strange quark matter, which would be present in the core of a quark or hybrid star, is expected to be in a superconducting state. One of the possible pairing patterns for strange matter in high-density cores is that of the color-flavor locked (CFL) state \cite{Negreiros-2012, Alford-2001}. In the CFL phase all quarks are paired. {Following the paradigm of several related studies in the literature~\cite{Negreiros-2012, Blaschke-2000, Blaschke-2001,Zapata-2022,Lyra-2023,Dexheimer-2013}, we effectively studied the impact of pairing among all quark flavors by introducing suppression factors to the relevant neutrino emission processes and the quark heat capacity. In particular,} the QDU processes are suppressed by a factor $e^{-\Delta_q/k_b T}$, while the QMU and QB processes suffer suppression by a factor $e^{-2\Delta_q/k_BT}$ for $T < T_{c_q}$. The parameter $\Delta_q$ represents the energy gap and $T_{c_q}$ is the critical temperature below which strange quark matter enters the superconducting state. The critical temperature for the CFL phase is believed to be smaller than the Bardeen--Cooper--Schrieffer (BCS) value $T_{c_q} \simeq 0.57\Delta_{0_q}/k_B$, and in the present work it was set equal to $T_{c_q} = 0.4\Delta_{0_q}/k_B$, as in references \cite{Negreiros-2012, Blaschke-2000, Blaschke-2001} ($\Delta_{0_q}$ corresponds to the gap at zero temperature). Furthermore, following Refs.~\cite{Blaschke-2001, Grigorian-2005}, we used the interpolation formula $\Delta_q(T) = \Delta_{0_q} \sqrt{1-T/T_{c_q}}$ to describe the temperature dependence of the gap. Finally, the presence of pairing phenomena affects the specific heat of the quarks by inserting the following suppression factor in \mbox{Equation (\ref{32})~\cite{Negreiros-2012, Blaschke-2000}} 
\begin{equation}\label{34}
c_{sq} \simeq 3.2\:\frac{T_{c_q}}{T}\left[2.5 - 1.7\frac{T}{T_{c_q}} + 3.6\left(\frac{T}{T_{c_q}}\right)^2  \right] e^{-\Delta_q/k_BT}.
\end{equation}

Notably, according to the work of Alford et al.~\cite{Alford-2001b,Alford-2005},
the condition that needs to be satisfied for the CFL state to be energetically favored is the following
\begin{equation}\label{67}
    \Delta_{0q}>\frac{3m_s^2}{2\mu_{_B}}.
\end{equation}
Given that the strange quark mass is considered to be 95 MeV and that the baryon chemical potential in quark matter should be at least 930 MeV (at least in the case of hybrid stars), one may conclude that $\Delta_{0_q}$ must be larger than $\sim$15 MeV. Therefore, this value was used in the scenario where the quark matter is considered to be in the CFL state. {It is worth mentioning that the exact selected value might not be of particular importance. More precisely, when using such large gaps, which permit the occurrence of a CFL like phase, the neutrino emission and quark heat capacity are severely suppressed (practically absent)~\cite{Weber-2005}.}

Notably, as mentioned in Ref.~\cite{Alford-2001b}, if the criterion of Equation~(\ref{67}) does not hold, a weaker form of pairing may still occur. For the latter reason, we included (similarly to previous studies, see Refs.~\cite{Blaschke-2000,Blaschke-2001,Negreiros-2012,Zapata-2022} and references therein) $\Delta_{0_q}$ values lower than \mbox{15 MeV}. In that case, $\Delta_{0_q}$ was treated essentially as a phenomenological parameter that does not correspond to the gap of an ideal CFL state, but encapsulates the effect of a weaker suppression. Following the work of Blaschke et al. \cite{Blaschke-2000}, in the case of low pairing gaps, the critical temperature was calculated via $T_{c_q} \simeq 0.57\Delta_{0_q}/k_B$.

Finally, we also considered the "2 color--flavor SuperConductor" (2SC) scenario \cite{Alford-2008}, in which only two colors of u and d quarks are paired (e.g. $\mathrm{u_g}$, $\mathrm{u_b}$, $\mathrm{d_g}$, $\mathrm{d_b}$), while the s quarks of all three different colors and the $\mathrm{u_r}$, $\mathrm{d_r}$ quarks remain unpaired. In that case, the contribution of the paired blue-green and green-blue u and d quarks in the total neutrino luminosity and the total heat capacity is suppressed due to large pairing gaps. However, the QDU, QMU, QB processes involving the unpaired s quarks and the red-colored u and d quarks are not suppressed. For instance, there is only one reaction which is not blocked by the large pairing gap out of the possible three of the d quark branch of the QDU process, which is $d_r \rightarrow u_r + e^- + \bar\nu_e$. Hence, the total neutrino emissivity (Equation~(\ref{19})) should be multiplied by the factor 1/3 \cite{Page-2002}. A similar analysis applies for the s quark branch of the QDU process. Moreover, Equation~(\ref{32}), which provides the specific heat capacity for all types and colors of unpaired quarks, should be scaled by a factor of 
9/5 \cite{Page-2002}. This factor accounts for five out of the total nine possible quark color--flavor combinations. The assumptions above are in good agreement with Blaschke's estimations \cite{Blaschke-2000, Yu-2006}, which dictate that the emissivities of all the neutrino emitting processes in the 2SC phase are reduced by approximately one order of magnitude. {A similar approach was considered} for the estimation of the suppression factors of the other two neutrino emitting processes (QB, QMU).

\subsection{Cooling Processes in Hadronic Matter}\label{subsec34}
The nucleon Direct Urca process (NDU) is the strongest neutrino emitting process in hadronic matter. It consists of two successive reactions, beta decay and capture \cite{Yakovlev-2001}
\begin{equation} \label{35}
    n \rightarrow p + e^- + \bar{\nu}_e,
\end{equation}
\begin{equation} \label{36}
p + e^- \rightarrow n + \nu_e.
\end{equation}
The total neutrino emissivities related to these processes are given by \cite{Yakovlev-2001, Yakovlev-1999} 
\begin{equation}\label{37}
\mathcal{E}_\nu^{NDU} \simeq 4.00\cdot10^{27}\left(\frac{n_e}{n_0} \right)^{1/3}\:\frac{m_n^*m_p^*}{m_n^2}\:T_9^6\:\Theta_{npe} \quad \mathrm{erg \cdot cm^{-3} \cdot s^{-1}},
\end{equation}
where $n_e$ is the electron number density, $m_j^*$ is the neutron's or proton's effective mass, which in the present work has been considered equal to $m_{j}^* = 0.7\cdot m_{j}$ with j = n, p. The $\Theta_{npe}$ factor is a step function which is equal to 1 if the Fermi momenta of the n, p and e satisfy the triangle condition $p_{F_n} < p_{F_p} + p_{F_e}$ \cite{Yakovlev-2001}. However, if the previous condition is not satisfied, the $\Theta_{npe}$ factor is equal to 0 and in this case the process is deactivated. In npe models the triangle condition is simplified and the NDU is activated if the proton fraction is higher than the critical value $x_c = 1/9$ \cite{Yakovlev-2001}.

In the case where the proton fraction does not exceed the value $x_c$, the dominant neutrino emission processes are the nucleon Modified Urca process (NMU) and the neutrino Bremsstrahlung process due to nucleon--nucleon scattering (NNB). The NMU process consists of two branches, the neutron and proton branch, which involve an additional nucleon spectator compared to NDU \cite{Yakovlev-2001}
\begin{equation} \label{38}
    n + n \rightarrow p + n + e^- + \bar{\nu}_e,
\end{equation}
\begin{equation} \label{39}
p + n + e^- \rightarrow n + n + \nu_e,
\end{equation}
\begin{equation} \label{40}
n + p \rightarrow p + p + e^- + \bar{\nu}_e,
\end{equation}
\begin{equation} \label{41}
p + p + e^- \rightarrow n + p + \nu_e.
\end{equation}
The emissivity of the neutron branch was calculated by Friman and Maxwell \cite{Friman-1979} and of the proton branch by Yakovlev and  Levenfish \cite{Yakovlev-1995}, and are, respectively, equal to
\begin{equation}\label{42}
\mathcal{E}_\nu^{NMU-n} \simeq 8.55\cdot10^{21}\left(\frac{m_n^*}{m_n} \right)^{3}\:\left(\frac{m_p^*}{m_p} \right)\:\left(\frac{n_e}{n_0} \right)^{1/3}\:T_9^8\:\alpha_n\:\beta_n \quad \mathrm{ erg \cdot cm^{-3} \cdot s^{-1}},
\end{equation}
\begin{equation}\label{43}
\mathcal{E}_\nu^{NMU-p} \simeq 8.53\cdot10^{21} \left(\frac{m_p^*}{m_p} \right)^3 \left(\frac{m_n^*}{m_n} \right) \left(\frac{n_e}{n_0}\right)^{1/3}\:T^8_9\:\alpha_p\:\beta_p\: \left(1 - \frac{p_{F_e}}{4p_{F_p}} \right)\: \Theta_{Mp} \quad \mathrm{ erg \cdot cm^{-3} \cdot s^{-1}},
\end{equation}
where $n_p$ is the proton number density and $p_{F_x}$ are the Fermi momenta of nucleons or electrons (x = n, p, e). Following Ref.~\cite{Yakovlev-2001}, the factors $\alpha_n$, $\beta_n$, $\alpha_p$, $\beta_p$ are set equal to $\alpha_n = \alpha_p = 1.13$ and $\beta_n = \beta_p = 0.68$. The expression of the proton branch of the NMU process includes the $\Theta_{Mp}$ factor, which, similarly to the NDU process, makes it a threshold reaction. In this particular case, the reaction is activated if the inequality $p_{F_n} < 3p_{F_p} + p_{F_e}$ is satisfied \cite{Yakovlev-2001}. In npe matter, the inequality is satisfied when the proton fraction exceeds the critical value $x_{cp} = 1/65$ \cite{Yakovlev-2001}. The specific condition is met anywhere inside the star's core. The Fermi momenta of the nucleons are
\begin{equation}\label{44}
p_{F_{x}} = \hbar \: (3\:\pi^2\:n_{x})^{1/3} \quad \mathrm{(x = n, p, e)}.
\end{equation}
The nucleon--nucleon scattering reactions and their respective emissivities are \cite{Yakovlev-2001}
\begin{equation} \label{45}
n + n \rightarrow n + n + \nu + \bar{\nu},
\end{equation}
\vspace{-1.9em}
\begin{equation} \label{46}
n + p \rightarrow n + p + \nu + \bar{\nu},
\end{equation}
\begin{equation} \label{47}
p + p \rightarrow p + p + \nu + \bar{\nu},
\end{equation}
\begin{equation}\label{48}
\mathcal{E}_\nu^{NNB-nn} \simeq 7.4\cdot10^{19}\left(\frac{m_n^*}{m_n} \right)^{4}\:\left(\frac{n_n}{n_0} \right)^{1/3}\:\alpha_{nn}\:\beta_{nn}\:N_{\nu}T_9^8 \quad \mathrm{ erg \cdot cm^{-3} \cdot s^{-1}},
\end{equation}
\begin{equation}\label{49}
\mathcal{E}_\nu^{NNB-np} \simeq 1.5\cdot10^{20}\left(\frac{m_n^*m_p^*}{m_n m_p}\right)^{2}\:\left(\frac{n_p}{n_0} \right)^{1/3}\:\alpha_{np}\:\beta_{np}\:N_{\nu}T_9^8 \quad \mathrm{ erg \cdot cm^{-3} \cdot s^{-1}},
\end{equation}
\begin{equation}\label{50}
\mathcal{E}_\nu^{NNB-pp} \simeq 7.4\cdot10^{19}\left(\frac{m_p^*}{m_p} \right)^{4}\:\left(\frac{n_p}{n_0} \right)^{1/3}\:\alpha_{pp}\:\beta_{pp}\:N_{\nu}T_9^8 \quad \mathrm{ erg \cdot cm^{-3} \cdot s^{-1}},
\end{equation}
where $\alpha_{NN}$ are dimensionless factors which similarly to Refs.~\cite{Yakovlev-2001, Yakovlev-1999} are taken to be $\alpha_{nn} = 0.59$, $\alpha_{np} = 1.06$, $\alpha_{pp} = 0.11$ and the correction factors $\beta_{NN}$ were set $\beta_{nn} = 0.56$, $\beta_{np} = 0.66$, $\beta_{pp} \approx 0.7$. Also, $N_\nu = 3$ is the number of the different neutrino flavors.

\subsection{Specific Heat in Hadronic Matter}\label{subsec35}
The total heat capacity in hadronic matter is equal to the sum of the contributions {from different} particle species (neutrons, protons and electrons are considered). The specific heat of neutrons and protons are \cite{Ofengeim-2016, Grigorian-2005a}
\begin{equation}\label{51}
c_{v_j} = \frac{k_B^2}{3\hbar^3}\:T\:m_j^*\:p_{F_j} \quad  \mathrm{(j = n, p)}.
\end{equation}
The electron's specific heat is given in Section \ref{subsec32}.

\subsection{Pairing in Hadronic Matter}\label{subsec36}
The presence of superfluidity plays a crucial role in the thermal evolution of hadronic matter in a neutron or a hybrid star. It is believed that superfluidity is of BCS type~\cite{Yakovlev-1998}. The neutron pairing occurs at subnuclear densities ($\rho \leq \rho_0 = 2.8 \cdot 10^{14} \;\mathrm{g \cdot cm^{-3}}$) due to nucleon--nucleon attraction in the $^1S_0$ state~\cite{Yakovlev-1998}. This range of densities corresponds to the inner crust of the star, which consists of nuclei, electrons, and superfluid neutrons of $^1S_0$ type. However, for densities greater than $\rho_0$, the singlet-state attraction of neutrons becomes repulsive and the triplet-state ($^3P_2$) attraction is favored leading to anisotropic gaps~\cite{Yakovlev-1998}. As a result, it is preferable for neutrons to form $^3P_2$ superfluids, due to extremely high densities. This range of densities is found in the star's core. Furthermore, the protons in the star's core form $^1S_0$ superfluids, due to the small number density compared to the one of neutrons, making the singlet-state pp interaction more attractive~\cite{Yakovlev-1998}. In the present work we consider the neutrons and protons of the core to form $^3P_2$ and $^1S_0$ superfluids, respectively. 

{Similarly} to the quark matter case, {pairing} in hadronic matter inserts suppression factors in the expressions of the neutrino emitting processes \cite{Grigorian-2005a}
\begin{equation}\label{52}
\xi_{jj} =
\begin{cases}
e^{-\Delta_{jj}/k_BT} & ,T < T_{c_j}\\[1ex]
1 & ,T \geq T_{c_j},
\end{cases}
\end{equation}
with j = n, p. $\Delta_{jj}$ is the superfluidity energy gap and $T_{c_j}$ is the critical temperature below which j particles form a superfluid \cite{Blaschke-2004, Grigorian-2005}. The reduction factors of each process are listed in Table~\ref{tab:table1}~ \cite{Grigorian-2005a}.
\begin{table}[h!] 
\caption{Reduction factors of each nucleon process~\cite{Grigorian-2005a}.\label{tab:table1}}
\newcolumntype{C}{>{\centering\arraybackslash}X}
\begin{tabularx}{\textwidth}{CC}
\toprule
\textbf{Process} & \textbf{Factor} \\
\midrule
NDU & $\min(\xi_{nn}, \xi_{pp})$ \\
NMU-n & $\xi_{nn} \cdot \min(\xi_{nn}, \xi_{pp})$ \\
NMU-p & $\xi_{pp} \cdot \min(\xi_{nn}, \xi_{pp})$ \\
NNB-nn & $\xi_{nn}^2$ \\
NNB-pp & $\xi_{pp}^2$ \\
NNB-np & $\xi_{nn} \cdot \min(\xi_{nn}, \xi_{pp})$ \\
\bottomrule
\end{tabularx}
\end{table}

The critical temperatures for each superfluidity case (isotropic, anisotropic) are approximately related to the superfluid energy gaps at zero temperature $\Delta_{0_{jj}} = \Delta_{jj}(0)$ via the following expressions \cite{Yakovlev-2001, Yakovlev-1999, Kaminker-2002}
\begin{equation}\label{53}
^1S_0 \; \mathrm{gap:}\; k_B\:T_{c_j} \simeq 0.5669\:\Delta_{0_{jj}},
\end{equation}
\begin{equation}\label{54}
^3P_2 \; \mathrm{gap:}\; k_B\:T_{c_j} \simeq 0.8416\:\Delta_{0_{jj}}.
\end{equation}
With respect to the zero temperature energy gap, we apply the following parametrization \cite{Andersson-2005, Kaminker-2001, Kaminker-2002, Ho-2015}
\begin{equation}\label{55}
\Delta_{0_{jj}}(p_{F_x}) = \Delta_0 \frac{(p_{F_x}-p_0)^2}{(p_{F_x}-p_0)^2+p_1} \frac{(p_{F_x}-p_2)^2}{(p_{F_x}-p_2)^2+p_3},
\end{equation}
where $\Delta_0$, $p_0$, $p_1$, $p_2$ and $p_3$ are fit parameters, which are determined by different superfluidity gap models. The interpolation formula between the zero temperature gap and the energy gap with respect to temperature is {of the form} $\Delta_{jj}(T) = \Delta_{0_{jj}} \sqrt{1-T/T_{c_j}}$ (see Ref.~\cite{Grigorian-2005} and references therein). The values of the aforementioned parameters that are being used in this study are presented in Table~\ref{tab:table2} (the considered numerical values are from Ref.~\cite{Andersson-2005}, where the authors utilized calculations for different gap models).

\begin{table}[h!] 
\caption{Superfluidity gap models for different superfluidity types. The numerical values are from Ref.~\cite{Andersson-2005}, where the authors utilized the calculations for different gap models from Refs.~\cite{Elgaroy-1996c,Amundsen-1985a,Baldo-1992,Baldo-1998} (appearing in the last column). The $\Delta_0$ parameter is measured in MeV, the $p_0$, $p_2$ parameters are measured in fm$^{-1}$ and the $p_1$, $p_3$ in fm$^{-2}$. In the pairing type column PS stands for proton singlet and NT for neutron triplet superfluidity type.\label{tab:table2}}
\newcolumntype{C}{>{\centering\arraybackslash}X}
\begin{tabularx}{\textwidth}{ccCCCCCC}
\toprule
\textbf{Pairing Type} & \textbf{Gap Model} & $\boldsymbol{\Delta_0}$ & $\boldsymbol{p_0}$ & $\boldsymbol{p_1}$ & $\boldsymbol{p_2}$ & $\boldsymbol{p_3}$ & \textbf{Ref.} \\
\midrule
\multirow{3}{*}{\shortstack{PS}} 
  & i & 61 & 0 & 6 & 1.1 & 0.6 & \cite{Elgaroy-1996c} \\
  & ii & 55 & 0.15 & 4 & 1.27 & 4 & \cite{Amundsen-1985a} \\
  & iii & 2.27 & 0.1 & 0.07 & 1.05 & 0.25 & \cite{Baldo-1992} \\
\midrule
\multirow{6}{*}{\shortstack{NT}} 
  & i & 4.8 & 1.07 & 1.8 & 3.2 & 2 & \cite{Baldo-1998} \\
  & ii & 10.2 & 1.09 & 3 & 3.45 & 2.5 & \cite{Baldo-1998} \\
  & iii & 2.2 & 1.05 & 1 & 2.82 & 0.6 & \cite{Baldo-1998} \\
  & iv & 0.425 & 1.1 & 0.5 & 2.7 & 0.5 & \cite{Elgaroy-1996b} \\
  & v & 0.068 & 1.28 & 0.1 & 2.37 & 0.02 & \cite{Elgaroy-1996b} \\
  & vi & 2.9 & 1.21 & 0.5 & 1.62 & 0.5 & \cite{Elgaroy-1996c} \\
\bottomrule
\end{tabularx}
\end{table}

It has to be noted that the previous formula in Equation~(\ref{55}) is valid when\linebreak $p_2< p_{F_x} < p_0$, while the energy gap vanishes for $p_{F_x} \leq p_0$ and $p_{F_x} \geq p_2$ \cite{Kaminker-2001}. The presence of superfluidity activates two additional neutrino emitting processes. These are the neutron pair breaking and formation (nPBF) and the proton pair breaking and formation (pPBF) \cite{Schaab-1997}. The emissivities of these processes are given by~\cite{Voskresensky-1987, Blaschke-2004} 
\begin{equation}\label{56}
\mathcal{E}_\nu^{jPBF} \sim 10^{29}\:m_{jPBF}^*\left[\frac{p_{F_j}(n_b)}{p_{F_n}(n_0)} \right]\:\left[\frac{\Delta_{jj}}{MeV} \right]^7\:\left[\frac{k_B\:T}{\Delta_{jj}} \right]^{1/2}\:\xi_{jj}^2\quad \mathrm{ erg \cdot cm^{-3} \cdot s^{-1}}, \quad T<T_{c_j},
\end{equation}
where $m_{jPBF} = m_j^*/m_N$ with $m_N$ being the nucleon's mass and j = n, p. As a result $m_{jPBF} \simeq x_{m_j}$ with $x_{m_j} = m_j^*/m_j$. 

The neutron and proton pairing phenomena affect the specific heats for each particle species. Consequently, the specific heat in Equation~(\ref{51}) has to be multiplied by $\xi_{nn}$ for neutrons and by $\xi_{pp}$ for protons \cite{Grigorian-2005a}.

\subsection{Crust and Envelope}\label{subsec37}
Following the work of Ref. \cite{Yakovlev-1999}, the crust contributes to the total neutrino emission via the Bremsstrahlung of electrons which scatter off atomic nuclei. The luminosity of this process was provided by Maxwell \cite{Maxwell-1979} with the following approximate formula
\begin{equation}
L_{br} = 1.65\cdot10^{39}\:\frac{M_{cr}}{M_\odot}\left(\frac{T_b}{10^9K}\right)^6e^{\nu_b} \quad erg\cdot s^{-1}
\end{equation}
where $M_{cr}$ is the mass of the crust, $T_b$ is the temperature at the crust--envelope transition point and $\nu_b$ is the value of the gravitational redshift at the same point. In addition, the contribution of the crust in the total heat capacity is being neglected as in Refs.~\cite{Yakovlev-1999,Blaschke-2001}.

The envelope, being the outermost layer of a compact star, plays a crucial role in the cooling process, since it acts as a thermal insulator. Several models for the envelope's chemical composition have been introduced, which attempt to describe the relation between the surface temperature ($T_s$) and the temperature in the crust--envelope boundary ($T_b$). In this work, we have considered two different models of chemical composition. Specifically, we consider an envelope consisting from heavier elements (iron-like) and an envelope made from lighter elements (helium-like). The relations between $T_b$ and $T_s$ for both cases are \cite{Gudmundsson-1983, Cumming-2017}
\begin{equation} \label{57}
T_{b_8} = 1.288\:(T_{s_6}^4/g_{s_{14}})^{0.455} \quad \mathrm{(Fe-envelope)},
\end{equation}
\begin{equation} \label{58}
T_{b_8} = 0.552\:(T_{s_6}^4/g_{s_{14}})^{0.413} \quad \mathrm{(He-envelope)},
\end{equation}
with $g_{s14} = \frac{GM}{10^{14}R^2} \left(1 - \frac{2GM}{Rc^2}\right)^{-1/2}$ and $M$, $R$ the mass and radius of the compact star, respectively. The relation between $T_b$ and $T_s$ affects significantly the photon emission from the star's surface. In particular, the redshifted photon luminosity is given by \cite{Yakovlev-2001} 
\begin{equation} \label{59}
L_{\gamma}^\infty = 4\pi \: R^2 \: \sigma \: T_s^4 \: e^{\nu(r_b)},
\end{equation}
where $e^{\nu(r)}$ is the time-time component of the metric tensor, $r_b$ is the radius corresponding to the envelope--crust transition point and $\sigma$ is the Stefan--Boltzmann constant. Notably, the envelope is a very thin layer of nucleonic matter and one can set $e^{\nu(r_b)/2} \approx e^{\nu(R)/2} = \sqrt{1 - 2GM/Rc^2}$ (following the work of Ref.~\cite{Yakovlev-2011}).

The chemical composition of heat-blanketing envelopes remains uncertain and depends on the formation and evolutionary history of the compact object. According to Ref.~\cite{Beznogov-2021}, earlier studies typically assumed that these envelopes form in very young, hot stars where light elements are consumed in thermonuclear reactions, leaving behind heavier, iron-like constituents. However, the composition can be significantly altered by additional processes, including fallback of material onto the compact object, the accretion of hydrogen and helium from a companion star \cite{Blaes-1992} and the nuclear burn out of the envelope's matter \cite{Beznogov-2021, Chiu-1964, Rosen-1968, Chang-2003, Wijngaarden-2019}. For further details on the subject we refer to the comprehensive work of Ref.~\cite{Beznogov-2021} (and references therein).

\subsection{Thermal Evolution of Compact Stars in the Isothermal Approximation}\label{subsec38}

The general relativistic equations describing the thermal evolution of a spherically symmetric star were derived by Thorne \cite{Thorne-1977}. However, if one considers an isothermal stellar interior, an assumption that is known to be reasonable a couple of centuries after a compact star's birth, then the cooling problem gets significantly simplified~\cite{Yakovlev-2001}. In the context of General Relativity, a star is isothermal if the redshifted temperature $T^\infty(t)=T(r,t)e^{\nu(r)/2}$ is constant throughout its interior. In that case, as shown by Glen and Sutherland~\cite{Glen-1980}, the cooling curves can be extracted by the solution of the global thermal balance equation~\cite{Glen-1980, Yakovlev-1999}
\begin{equation}\label{60}
C(T^\infty)\frac{dT^\infty}{dt} = -L_{\nu}^\infty(T^\infty) - L_{\gamma}^\infty(T_s),
\end{equation}
\begin{equation}\label{61}
C = \sum_{i}\int c_{vi}\:dV, 
\end{equation}
\begin{equation}\label{62}
 L_{\nu}^\infty = \sum_{i}\int \mathcal{E}_i\:e^{\nu(r)}dV,
\end{equation}
\begin{equation}\label{63}
dV = 4\:\pi\:r^2\:\left[1 - \frac{2\:G\:m(r)}{rc^2}\right]^{-1/2}dr,
\end{equation}
where $C$ is the total heat capacity, $L_{\nu}^\infty$ corresponds to the total redshifted neutrino luminosity, and $m(r)$ is the gravitational mass distribution of the star. In Equation~(\ref{61}) the index $i$ denotes different particle species (neutrons, protons, electrons, and quarks), while in Equation~(\ref{62}) it is related to the different neutrino emitting processes (QDU, QMU, QB, NDU, NMU, NNB, PBF).

The photon luminosity in Equation~(\ref{59}) is expressed with respect to $T_s$. Therefore, in order to solve the aforementioned differential equation (Equation~(\ref{60})), the surface temperature must be transformed into the internal redshifted temperature $T^\infty$. The internal redshifted temperature in the crust--envelope boundary ($T_{b_8}^\infty$) is
\begin{equation}\label{64}
T_{b_8}^\infty = T_{b8} \left(1 - \frac{2GM}{c^2R}\right)^{1/2}.
\end{equation}
If we solve the systems of Equations~(\ref{64}), (\ref{57}) and (\ref{64}), (\ref{58}) with respect to $T_s$ we get
\begin{equation}\label{65}
T_{s-Fe} = 0.87 \cdot 10^6 \left[\frac{GM}{10^{14}R^2} \left( 1 - \frac{2GM}{Rc^2} \right)^{-1.6} (T_{b_8-Fe}^\infty)^{2.2} \right]^{1/4},
\end{equation}
\begin{equation}\label{66}
T_{s-He} = 1.43 \cdot 10^6 \left[\frac{GM}{10^{14}R^2} \left( 1 - \frac{2GM}{Rc^2} \right)^{-1.7} (T_{b_8-He}^\infty)^{2.4} \right]^{1/4}.
\end{equation}
In the present work, we study the evolution of the redshifted surface temperature, which is given by $T_{s-env}^\infty = T_{s-env} \sqrt{1 - 2GM/Rc^2}$. The label env corresponds to the previous chemical composition models (env = Fe, He).

\section{Results and Discussion}\label{sec4}

In the present work we considered four different EOSs. In particular, we have constructed one purely hadronic EOS, one purely quark EOS and two hybrid models. The low mass and radius of the CCO in the HESS J1731-347 remnant did not allow for a variety of different parametrizations for the purely hadronic EOS. More precisely, one has to use a rather low value for the symmetry energy slope and the incompressibility to achieve the small radius. On the contrary, in the case of hybrid models, the emergence of twin star solutions allowed us to consider two distinct parametrizations for the stiffness of the low density phase. The first parametrization is based on the comprehensive analysis of Lattimer~\cite{Lattimer-2023}, which involved numerous different works on the nuclear saturation properties. The second parametrization is based on the well-known PREX-II experiment~\cite{Adhikari-2021,Reed-2021}. As a consequence, the second hybrid EOS enables the activation of the NDU process due to its extreme stiffness. The exact parameter values employed in this study can be found in Table~\ref{tab:table3}. {Note that the parameter $K_{sym}$ is currently poorly constrained. Therefore, we have selected values that are compatible to the predictions of traditional nuclear models~\cite{Raduta-2018,Tews-2017,Gil-2022}}. It is worth mentioning that the softer parametrizations considered in this study (hadronic model and the first hybrid model) are also in accordance with $\chi$EFT calculations (at least at a conservative momentum scale of $k=p_F$, see Table~\ref{tab:table_new})~\cite{Tews-2018}. On the contrary, the PREX-II results are known to be in tension with other nuclear physics data, a discrepancy that has not been fully reconciled up to this moment. In any case, given that the PREX-II estimation opens a new cooling scenario for HESS J1731-347 (NDU activation) it is particularly important to include its potential implications in the CCO's thermal evolution.

\begin{table}[h!] 
\caption{Parametrizations for the EOSs constructed in the present work. Note that in the case of hybrid and quark EOSs the parameter $\beta$ is fixed to 0.1.\label{tab:table3}}
\newcolumntype{C}{>{\centering\arraybackslash}X}
\begin{tabularx}{\textwidth}{cCCCCCCC}
\toprule
\textbf{EOS Model} & \boldmath{$J$} \textbf{(MeV)} & \boldmath{$L$} \textbf{(MeV)} & \boldmath{$K_0$} \textbf{(MeV)}  &  \boldmath{$K_{sym}$} \textbf{(MeV)}  & \boldmath{$B_{as}^{1/4}$} \textbf{(MeV)}  & \boldmath{$B_{0}^{1/4}$} \textbf{(MeV)}  & \boldmath{$G_v$} \textbf{(fm}\boldmath{$^2$}\textbf{)}  \\
\midrule
Quark & - & - & - & - & 140 & 140  & 0.25 \\
Hadronic & 32 & 40 & 200 & $-$35 & - & - & - \\
Hybrid-1  & 32 & 52 & 230 & $-$100 & 57 & 185 & 0.28 \\
Hybrid-2  & 38 & 106 & 240 & 0 & 57 & 190 & 0.27  \\
\bottomrule
\end{tabularx}
\end{table}

\begin{table}[h!] 
\caption{Predictions of the soft hadronic EOSs constructed in this study vs $\chi$EFT predictions~\cite{Tews-2018}. The following predictions are for pure neutron matter.}
\label{tab:table_new}
\newcolumntype{C}{>{\centering\arraybackslash}X}
\begin{tabularx}{\textwidth}{CCCCC}
\toprule
 & \textbf{\boldmath $n_b/n_0$} & \boldmath{$\chi$}\textbf{EFT}  & \textbf{Hadronic} & \textbf{Hybrid-1} \\
\midrule
\multirow{2}{*}{$E_b$ (MeV)} 
  & 1 & 17.3 $\pm$ 3.8  & 16   & 16   \\
  & 2 & 36.9 $\pm$ 16.9 & 38.5 & 40.6 \\
\midrule
\multirow{2}{*}{$P$ (MeV $\cdot$ fm$^{-3}$)} 
  & 1 & 2.4 $\pm$ 0.6 & 2.1 & 2.7 \\
  & 2 & 15.1 $\pm$ 4.7 & 19.6 & 19.7 \\
\bottomrule
\end{tabularx}
\end{table}

\begin{figure}[h!]
\hspace{-0.7cm}
    \includegraphics[width=\textwidth]{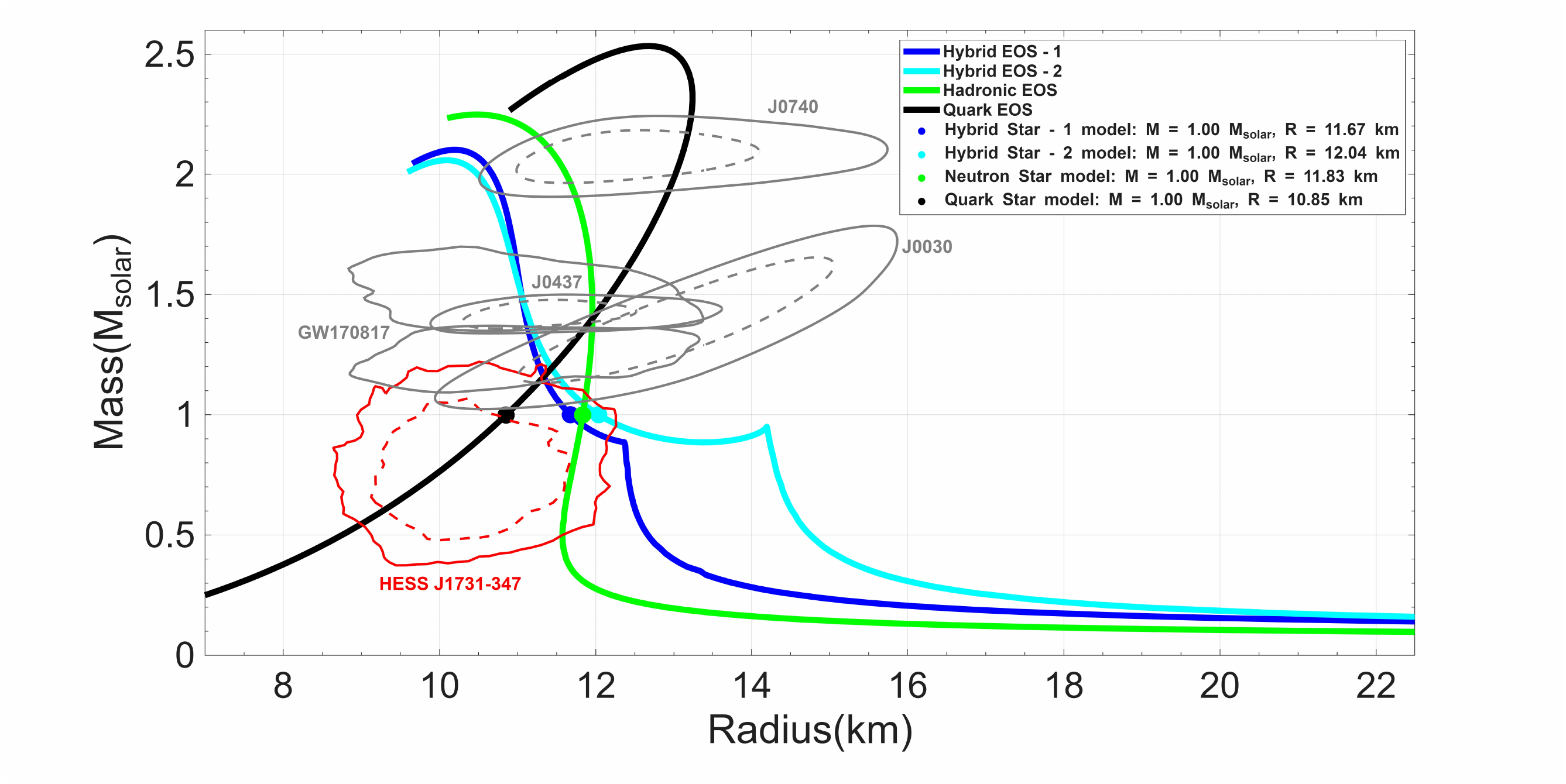}
    \caption{Mass–radius curves for each of the constructed EOSs. The black solid line corresponds to the quark EOS, the green line to the hadronic EOS, and the blue and cyan curves to the hybrid EOSs. The dots correspond to the compact star configurations listed in Table~\ref{tab:table4}. The gray contour regions denote mass and radius measurements with respect to PSR J0740+6620 \cite{Salmi-2024}, PSR J0030+0451 \cite{Miller-2019}, PSR J0437-4715 \cite{Choudhury-2024}, GW170817 \cite{Abbott-2018}. The red contour region denotes the mass and radius measurement of the CCO in HESS J1731-347 \cite{Doroshenko-2022}. The solid contours correspond to the $2\sigma$ confidence, while the dashed contours to the $1\sigma$ confidence.}
    \label{fig:fig1}
\end{figure}

In Figure~\ref{fig:fig1} one can find the mass--radius dependence predicted by the constructed EOSs. The scattered circular point appearing on each curve denotes the configuration for which the thermal evolution was studied. The structural properties of these configurations can be found in Table~\ref{tab:table4}. Notably, we found that all of the constructed EOSs are in accordance with state-of-the-art astronomical constraints from the LIGO/Virgo collaboration (GW170817 \cite{Abbott-2018}) and the NICER mission (PSR J0740+6620 \cite{Salmi-2024}, PSR J0030+0451 \cite{Miller-2019}, PSR J0437-4715 \cite{Choudhury-2024}). Despite the fact that the employed hadronic model can reconcile all of the modern constraints, it is crucial to clarify that its behavior at high densities is not on strong theoretical ground. This is related to the fact that it derives from an extrapolation of the energy per nucleon expansion around the saturation density and, therefore, it may lead to inaccurate results for high densities.

\begin{table}[h!] 
\caption{Masses, radii, and central baryon densities of the compact star models constructed in the present work.\label{tab:table4}}
\newcolumntype{C}{>{\centering\arraybackslash}X}
\begin{tabularx}{\textwidth}{cCCC}
\toprule
\textbf{Compact Star Model} & \boldmath{$M (M_\odot$)} & \boldmath{$R$ (km)} & {\boldmath $n_{b_{c}}/n_0$} \\
\midrule
Quark Star (QS) & 1.00 & 10.85 & 1.71  \\
Neutron Star (NS) & 1.00 & 11.83 & 2.39\\
Hybrid Star-1 (HS1) & 1.00 & 11.67 & 3.49 \\
Hybrid Star-2 (HS2) & 1.00 & 12.04 & 3.69\\
\bottomrule
\end{tabularx}
\end{table}

\subsection{Neutron Stars}
\begin{figure}[h!]
        \includegraphics[width=0.8\textwidth]{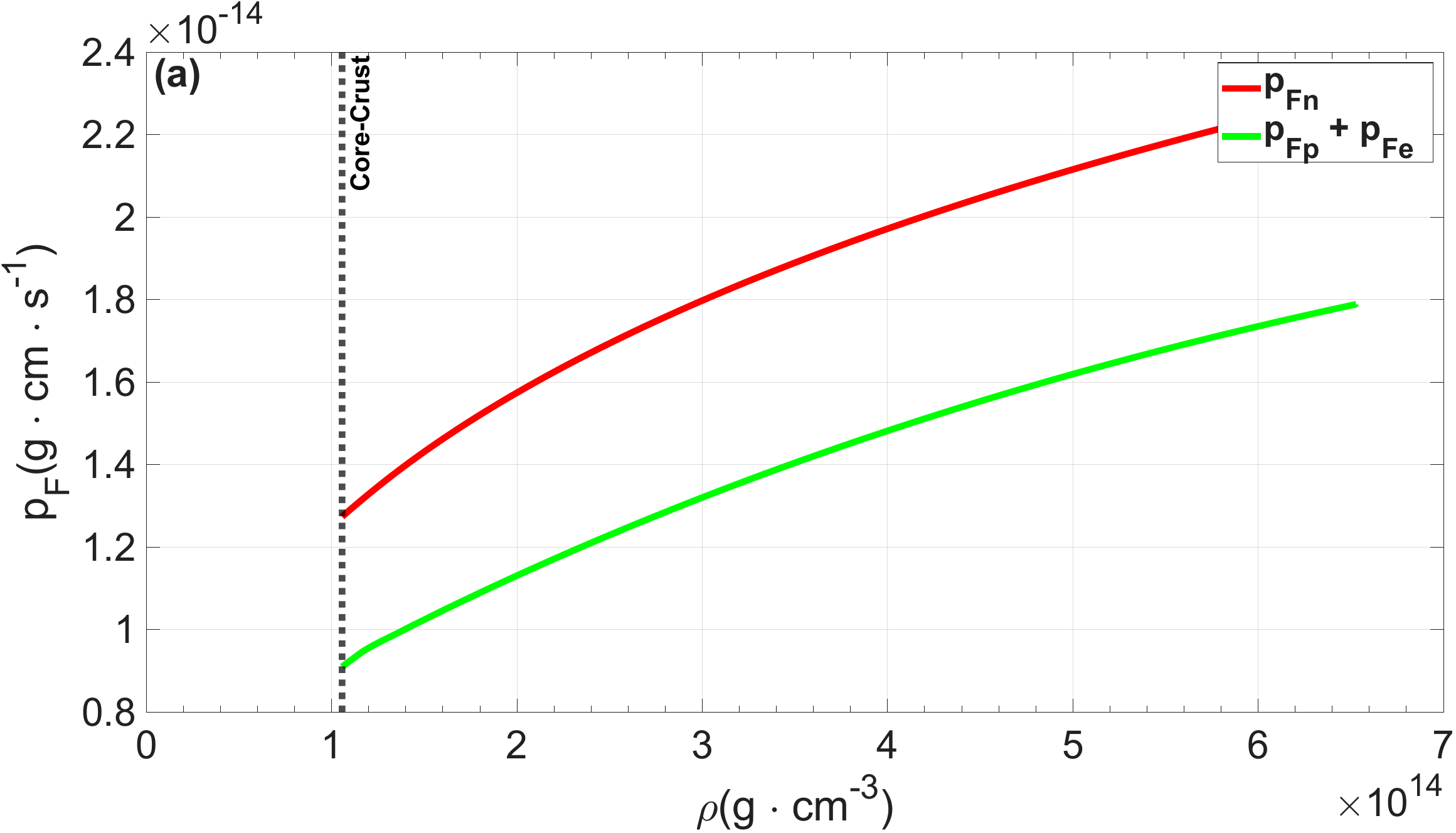} 
        \includegraphics[width=0.8\textwidth]{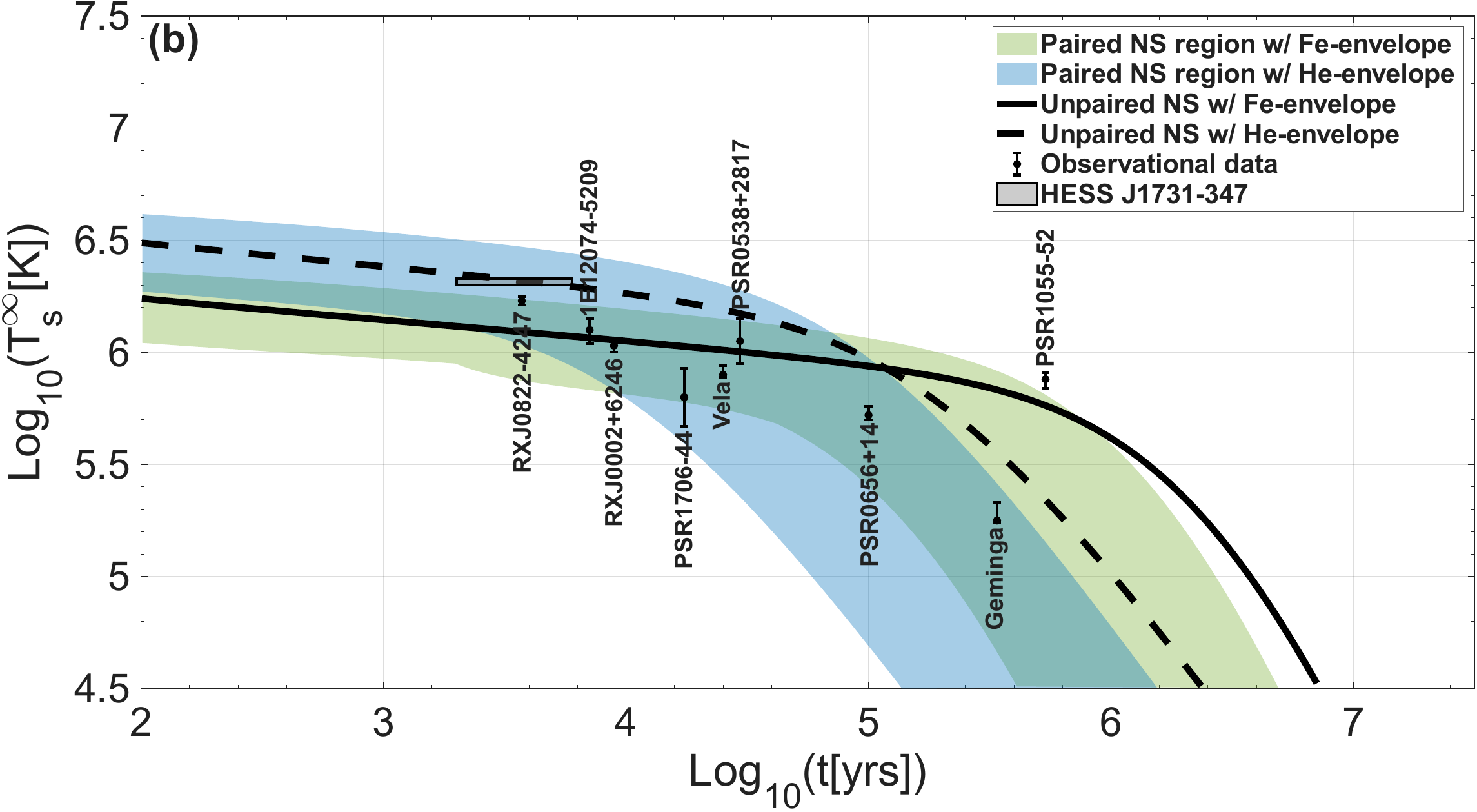} 
    \caption{(\textbf{a}) Fermi momenta of different particle species as a function of density for the NS model. The red curve corresponds to the neutrons Fermi momentum and the green curve to the sum of the Fermi momenta for protons and electrons.
    (\textbf{b}) Evolution of the redshifted surface temperature for the NS model considering different superfluidity models. The blue region corresponds to NSs with He-envelopes and the respective green region to NSs with Fe-envelopes. The solid line denotes the case of a NS with Fe-envelope where no pairing effects were considered, while the dashed line the case of a non superfluid NS with He-envelope. The gray shaded rectangle corresponds to the estimated surface temperature and age of the HESS J1731-347.} 
    \label{fig:fig2}
\end{figure}

Let us now proceed to the results for the thermal evolution of the NS model. We performed 18 cooling simulations, each corresponding to a different combination of PS and NT superfluidity models from Table~\ref{tab:table2}. As a result, we extracted the same number of cooling curves and determined the evolution of the redshifted surface temperature. This process was carried out for both scenarios related to the envelope's chemical composition (see Section~\ref{subsec37}). In Figure~\ref{fig:fig2}a, we present the Fermi momentum of neutrons and the sum of Fermi momenta for protons and electrons as a function of density. As it is clear, the triangle condition is not satisfied and hence the NDU process is not activated. 

In Figure~\ref{fig:fig2}b we present the resulting cooling curves. The light blue region corresponds to the thermal evolution of superfluid NSs with envelopes composed of light elements (He) and the light green region represents the cooling of superfluid NSs with heavy-element envelopes (Fe). In addition, the black dashed and solid curves correspond to the NS cases with envelopes composed by helium and iron-like elements without, however, considering the effects of neutron or proton superfluidity. The data points in \mbox{Figures~\ref{fig:fig2}b,~\ref{fig:fig3}b,~\ref{fig:fig4}b,~\ref{fig:fig5}b,c, and~\ref{fig:fig6}a,b} related to the observation of several pulsars were collected by the works of Refs. \cite{Yakovlev-1999, Yakovlev-2001, Page-2004} (see also references therein).

As shown in Figure~\ref{fig:fig2}b, there are two possible cases that seem to reconcile the surface temperature and age estimation for the CCO in the HESS J1731-347 remnant. Those are both the superfluid and non superfluid NS models with envelopes constructed by lighter elements. Interestingly, these results differ from previous works~\cite{Sagun-2023,Zhang-2025}, which have indicated that the presence of superfluidity may be essential for the explanation of the HESS J1731-347 constraints in the framework of hadronic stars. In the case of a heavy element envelope we found that the superfluid NS models fail to account for the CCO's properties, while the non superfluid model exhibits significant deviations from the corresponding temperature-age data. Notably, our results for the non superfluid NS cases are in excellent agreement with the seminal paper of Ref.~\cite{Page-2004} (see Figure~16). 

It is worth noting that the explanation of the HESS J1731-347 constraints crucially depends on the non activation of the NDU process. The absence of NDU driven cooling is expected for a NS of such low mass and radius since such structural properties favor soft nuclear models. Interestingly, soft parametrizations (predicting low radii and tidal deformabilities) are known to have high mass thresholds for the activation of the NDU process~\cite{Lopes-2024,Scurto-2025,Sarkar-2023}.

It is important to clarify that it is yet not clear how such a low mass NS could be formed. In particular, the comprehensive analysis of Ref.~\cite{Suwa-2018} suggests that NSs with masses lower than 1.17 $\mathrm{M_{\odot}}$ might not occur at the event of a supernova explosion. Interestingly, the recent work of Ref.~\cite{Zhang-2025} has proposed a possible astrophysical path for the formation of such a low mass NS.

\subsection{Hybrid Stars}

\begin{figure}[h!]

         \includegraphics[width=0.8\textwidth]{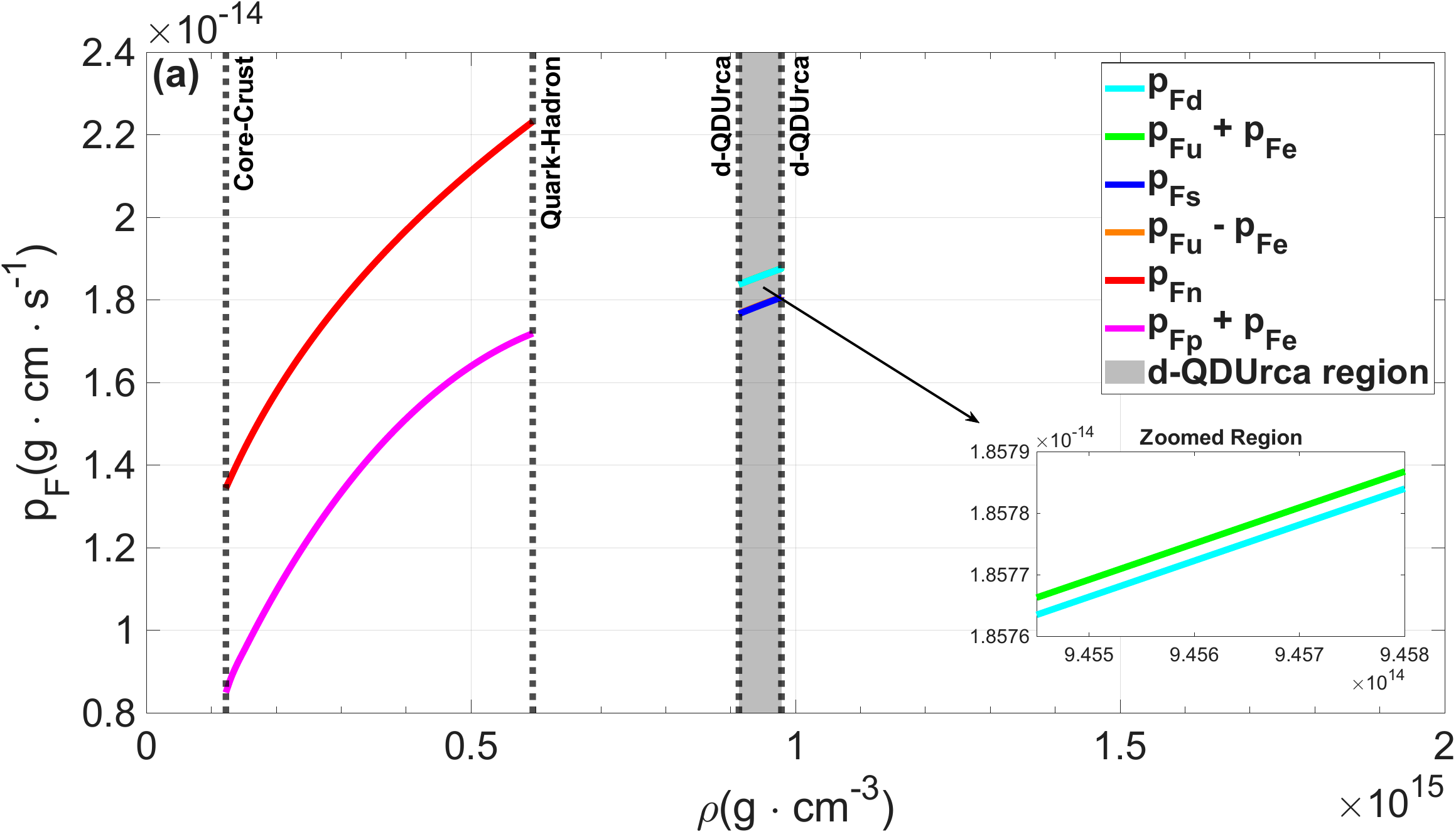}
        \includegraphics[width=0.8\textwidth]{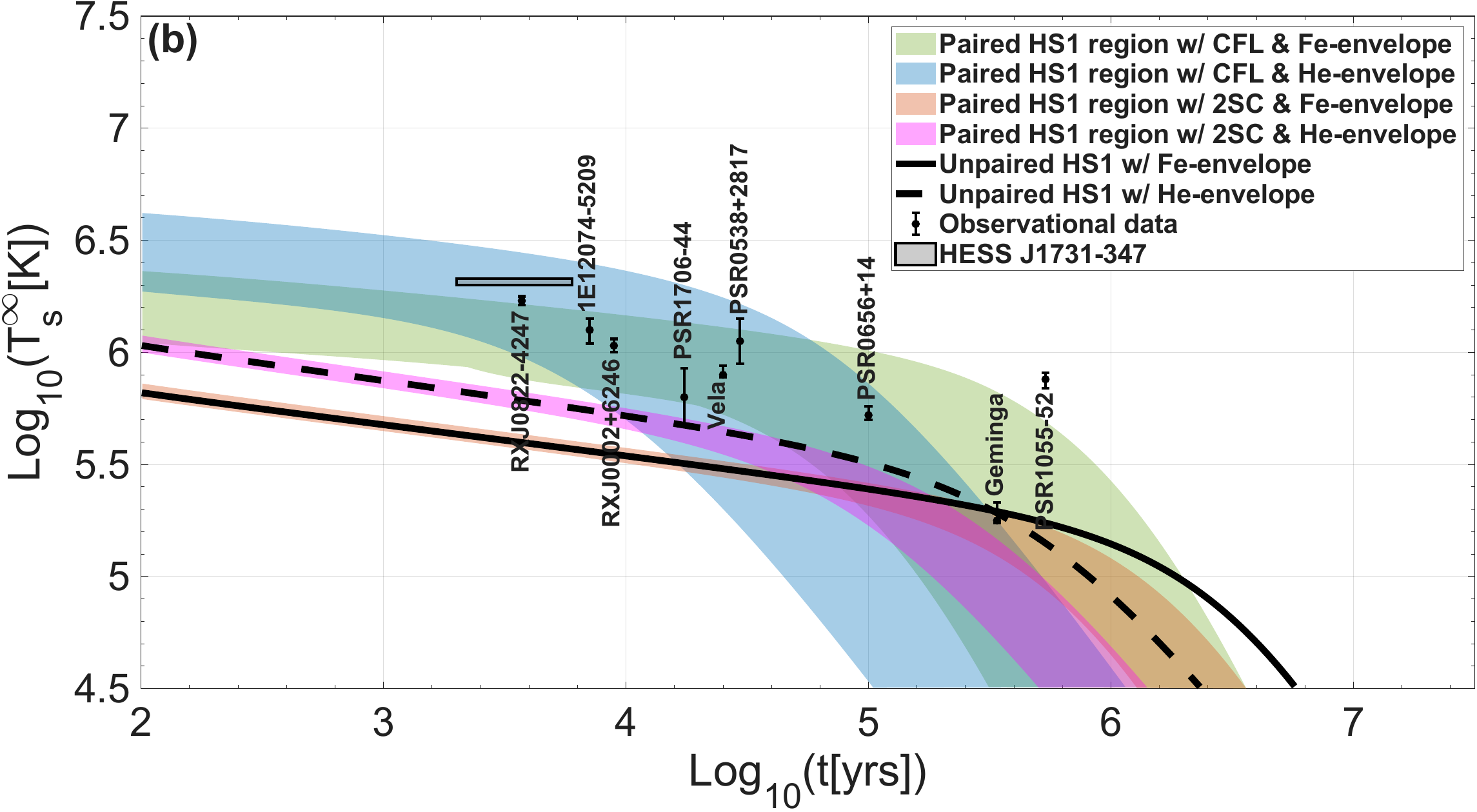}

    \caption{(\textbf{a}) Fermi momenta of different particle species as a function of density for the HS1 model. The red solid line corresponds to the Fermi momentum of neutrons, the magenta solid line to the sum of the Fermi momenta for protons and electrons, the cyan solid line to the Fermi momentum of d quarks, the blue solid line to the Fermi momentum of s quarks, the green solid line to the sum of the Fermi momenta for u quarks and electrons and the orange solid line to the difference of the Fermi momenta u quarks and electrons. The gray shaded region corresponds to the part of the star where the d-QDU is activated.
    (\textbf{b}) Evolution of the redshifted surface temperature with respect to time for the HS1 model. The blue region corresponds to a HS which suffers CFL superconductivity with $\Delta_{0_q} = 15$ MeV and has a light-element envelope, while the green region to a HS suffering the same pairing pattern and having a heavy-element envelope. The orange and the pink shaded regions correspond to HSs which suffer 2SC superconductivity. The orange (pink) region corresponds to models with an envelope of iron-like (helium-like) elements. The black solid and dashed lines describe the thermal evolution of unpaired HS models with different envelope composition. The gray shaded rectangle corresponds to the estimated surface temperature and age of the HESS J1731-347.}
    \label{fig:fig3}
\end{figure}

\begin{figure}[h!]

        \includegraphics[width=0.8\textwidth]{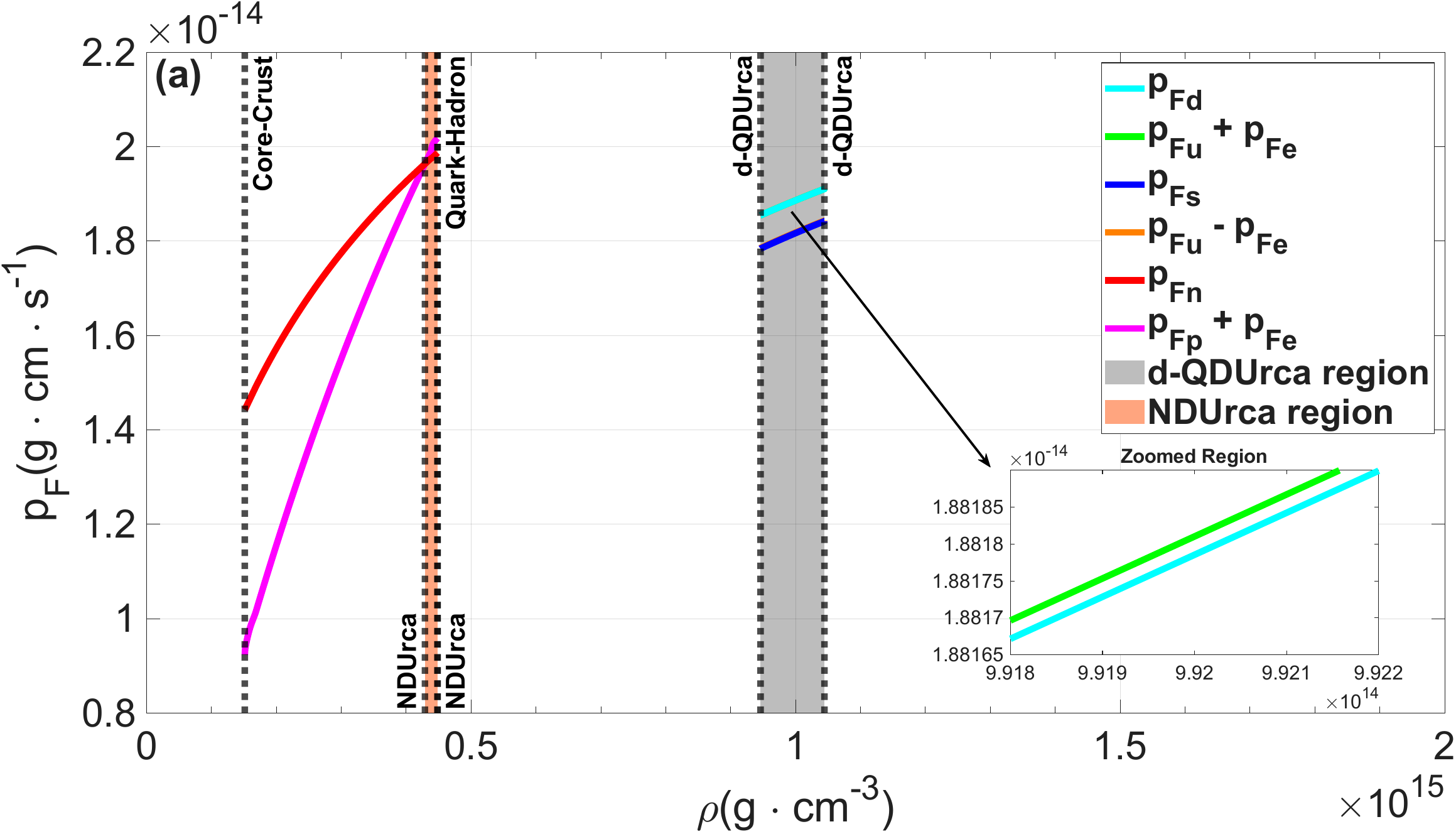}
        \includegraphics[width=0.8\textwidth]{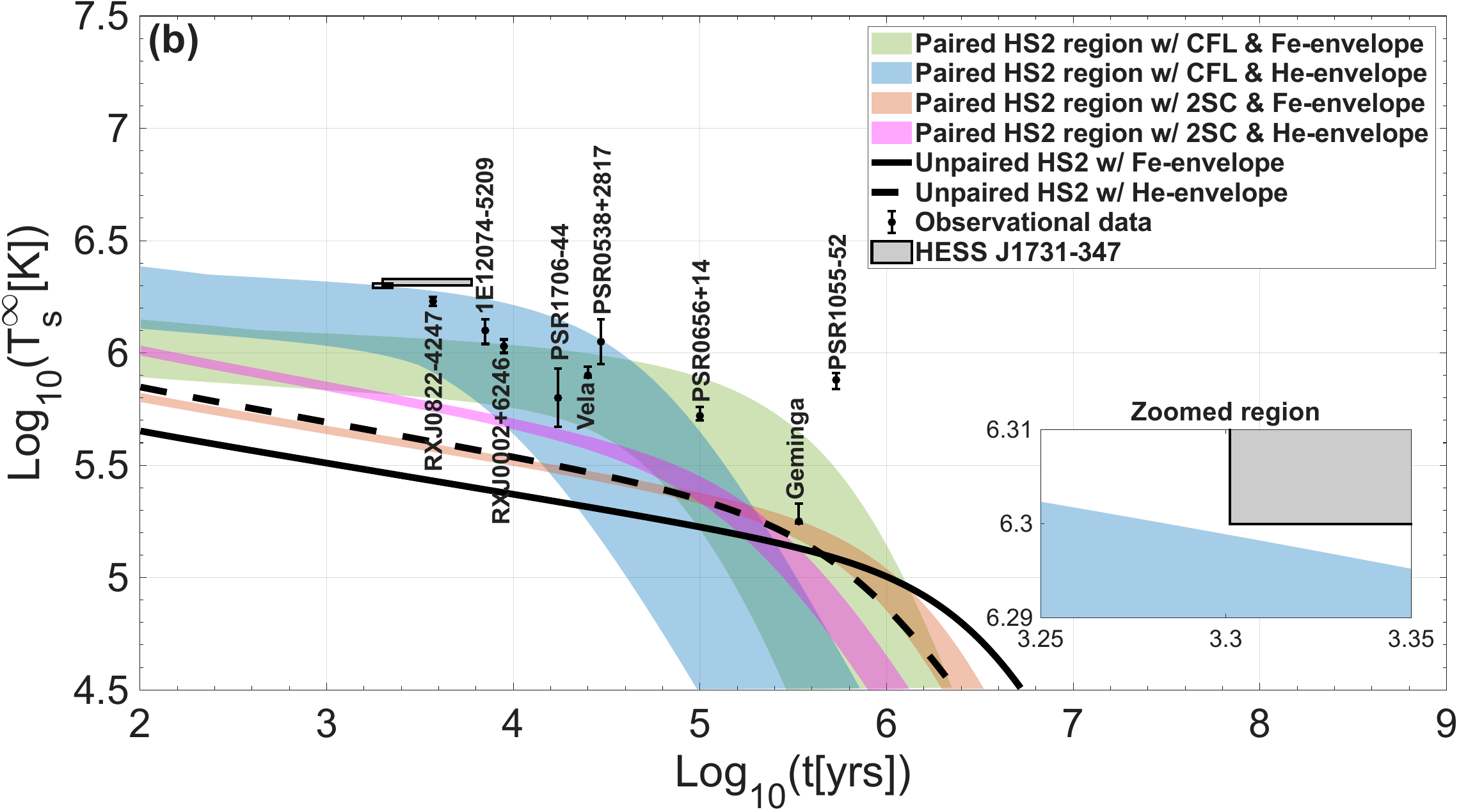}

    \caption{(\textbf{a}) Fermi momenta versus density for HS2. The colors of the solid lines and the gray shaded region correspond to the same parameters as in Figure~\ref{fig:fig3}a. The orange shaded region corresponds to the density area where the NDU is activated.
    (\textbf{b}) Evolution of the redshifted surface temperature with respect to time for the HS2 model. The colors of the shaded regions and the style of the curves correspond to the same cases of nucleonic envelopes and pairing types as in Figure~\ref{fig:fig3}b.}
    \label{fig:fig4}
\end{figure}

In the case where the HESS J1731-347 CCO was thought to be a hybrid compact star, we considered two different hybrid models. The first model, which is denoted as Hybrid Star-1 (HS1) in Table~\ref{tab:table4}, is described by a cooling scenario where the d quark branch of the QDU process is on, while the s quark branch of the QDU and the NDU process are not activated. This is clearly shown in Figure~\ref{fig:fig3}a, where in the hadronic region of the stellar configuration the sum of the Fermi momenta for protons and electrons is less than the Fermi momentum of neutrons. In the quark region, the Fermi momenta of the s and d quarks are less than the sum of the electrons' and u quarks' Fermi momenta. However, the difference between the Fermi momenta for the u quarks and the electrons is greater than the Fermi momentum of the s quark. Hence only the d-QDU is activated. 

Similarly to the case of the NS, we performed numerous different cooling simulations, one for each combination of proton and neutron superfluidity model from Table~\ref{tab:table2}. In addition, we considered both cases of quark superconductivity explained in Section~\ref{subsec33} (CFL and 2SC). Figure~\ref{fig:fig3}b illustrates the thermal evolution of HS1 for all the different combinations of superfluidity--superconductivity models considering both cases concerning the envelope's chemical composition. The black solid and dashed lines, which depict the cases where no pairing was considered, deviate significantly from the HESS J1731-347 constraints. This is related to the activation of the strong QDU process, which leads to rapid cooling. Notably, we found that the thermal evolution of the paired HS with a light-element envelope and CFL like pairing in its core could reproduce the temperature and age for the CCO in the HESS J1731-347 remnant. This was not the case when the HS1 model was studied under the assumption of a heavy element envelope (regardless of pairing effects), as well as for the scenario with 2SC like superconductivity and an envelope constructed by helium-like elements.

The thermal evolution of the second HS model, denoted as Hybrid Star-2 (HS2) in Table~\ref{tab:table3}, was governed by the d-QDU and the NDU processes. This is shown in Figure~\ref{fig:fig4}a, where the Fermi momenta of the s and d quarks are less than the sum of the Fermi momenta for u quarks and electrons throughout the entire quark region, while the s quark Fermi momentum is less than the difference of the u quarks and electrons Fermi momenta. In addition, the Fermi momenta of nucleons and electrons satisfy the triangle inequality in a part of the hadronic region of the stellar configuration.   

Similarly to the case of the HS1 model, we performed cooling simulations under the same superfluidity and superconductivity conditions as before. We also considered both instances of the envelope's chemical composition. The results are presented in Figure~\ref{fig:fig4}b. The activation of the NDU process enhances the reduction of the surface temperature compared to the previously studied models.  

As a consequence, the cooling curves which correspond to the non superfluid/ superconductive simulations fail to account for the HESS J1731-347 constraints. In addition, the present model slightly deviates from the desired contour region (in Figure~\ref{fig:fig4}b) even when pairing effects are included, with the deviations being larger in the 2SC scenario. In conclusion, the activation of the strong NDU process, even within a very small part of the hadronic region, makes this model a less compelling candidate for describing the thermal state of HESS J1731-347. The latter highlights the role of the symmetry energy parametrization of the low density phase. In particular, parametrizations of hadronic matter that are stiff enough to enable the NDU process may not be in accordance to the HESS J1731-347 data.

\subsection{Strange Quark Stars}

\begin{figure}[h!]

        \includegraphics[width=0.8\textwidth]{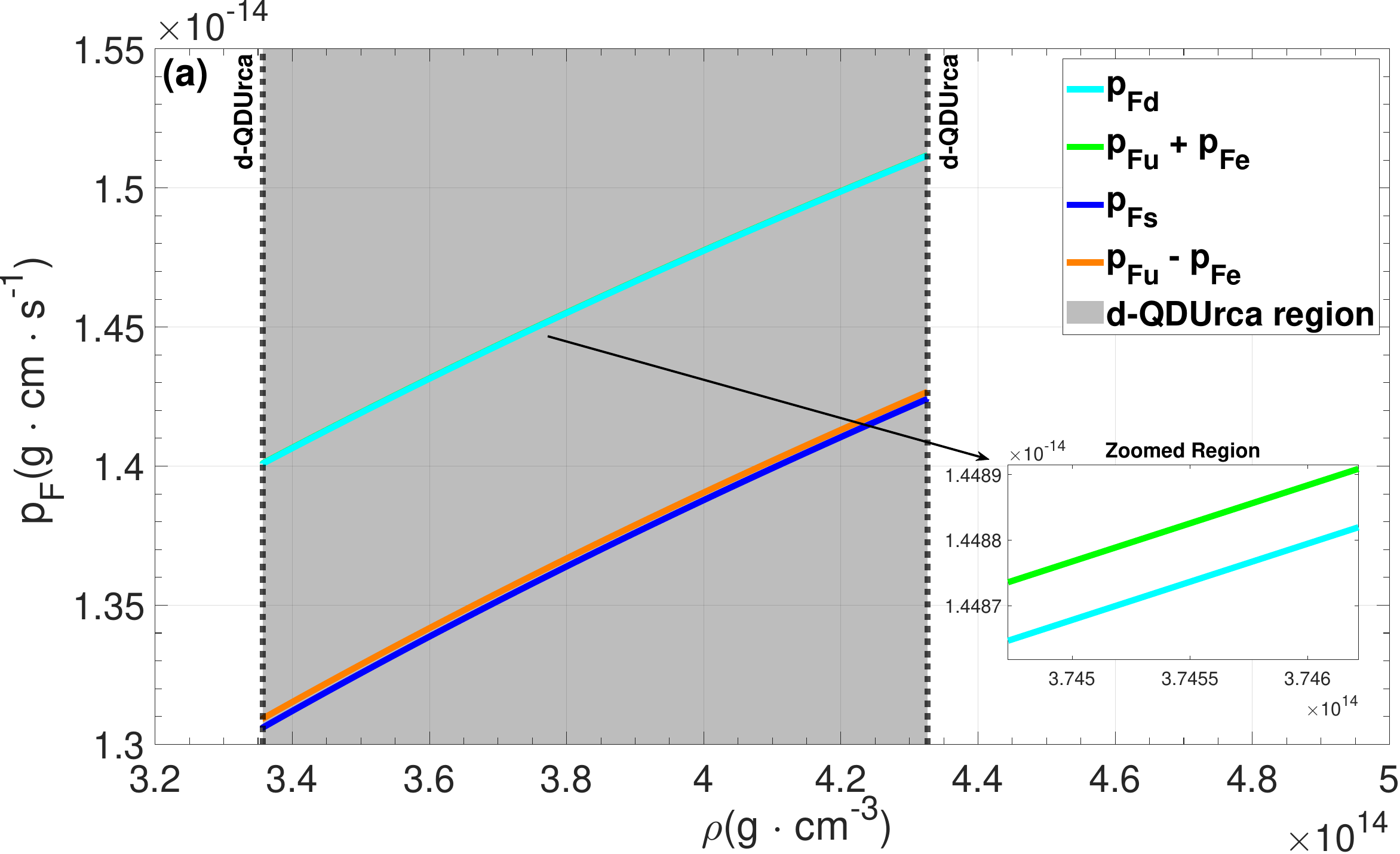}

\hfill

        \includegraphics[width=0.8\textwidth]{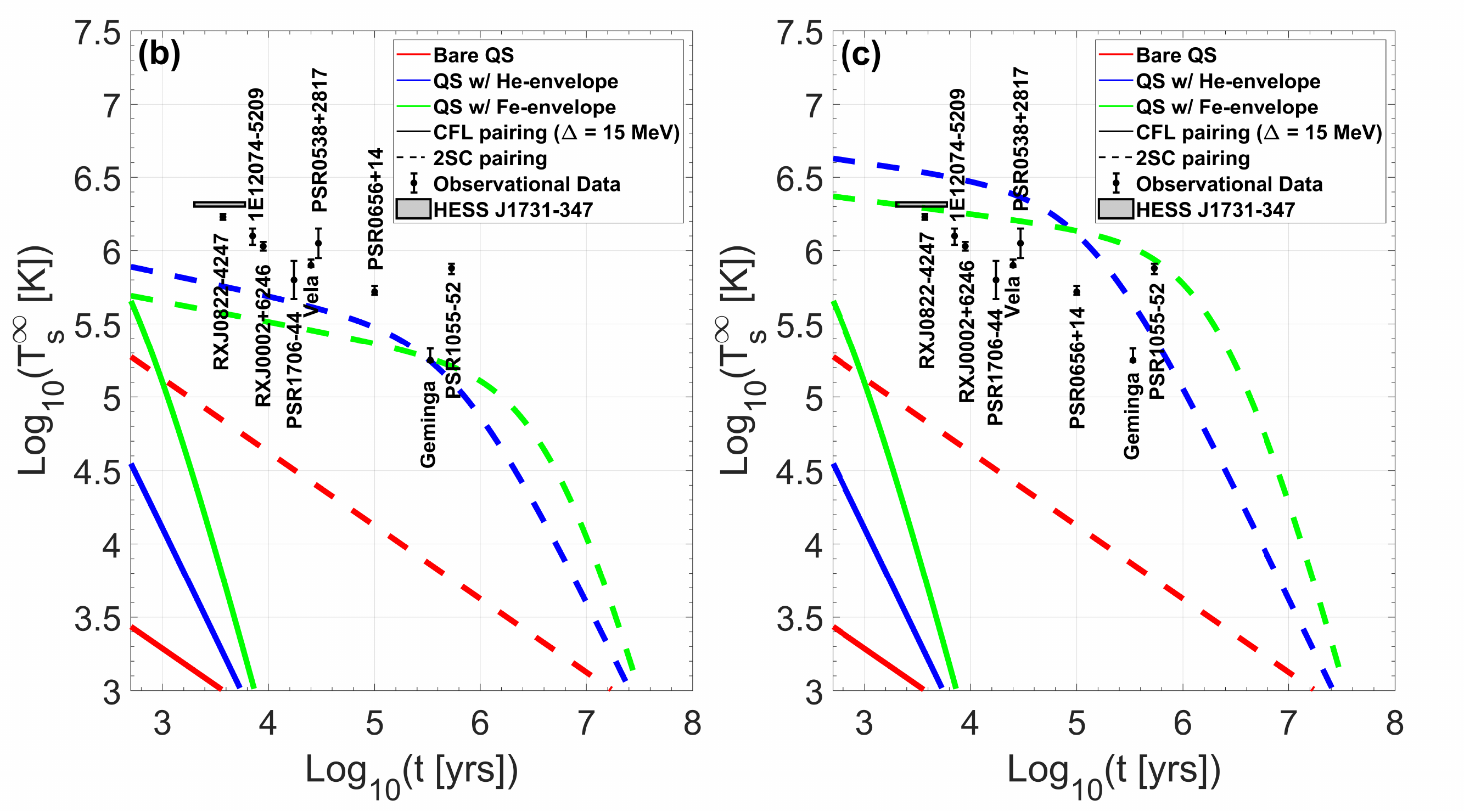}

    \caption{(\textbf{a}) Fermi momenta as a function of density for the QS model. The cyan solid line corresponds to the d quark Fermi momentum, the blue solid line to the s quark Fermi momentum, the green solid line to the sum of the u quark and electron Fermi momenta and the orange solid line to the difference between the u quarks and electrons Fermi momenta. These predictions were derived by considering the quark masses of Section~\ref{sec2}. The gray shaded region corresponds to the density range where the d-QDU is activated which is equal to the total size of the QS.
    (\textbf{b}) Redshifted surface temperature as a function of time for the QS model, in the case where the quark masses are those of Section~\ref{sec2} (predicting the Fermi momenta of Figure~\ref{fig:fig5}a). 
    (\textbf{c}) Redshifted surface temperature as a function of time for the QS model, considering a combination of quark masses that blocks the d-QDU process (see Appendix~\ref{app:A}). 
    In both panels (\textbf{b},\textbf{c}), solid (dashed) lines denote the case of CFL (2SC) pairing. Different colors denote different envelope composition (see legend). In addition, the gray shaded rectangle corresponds to the estimated surface temperature and age of the HESS J1731-347.}
    \label{fig:fig5}
\end{figure}

\begin{figure}[h!]

            \includegraphics[width=0.8\textwidth]{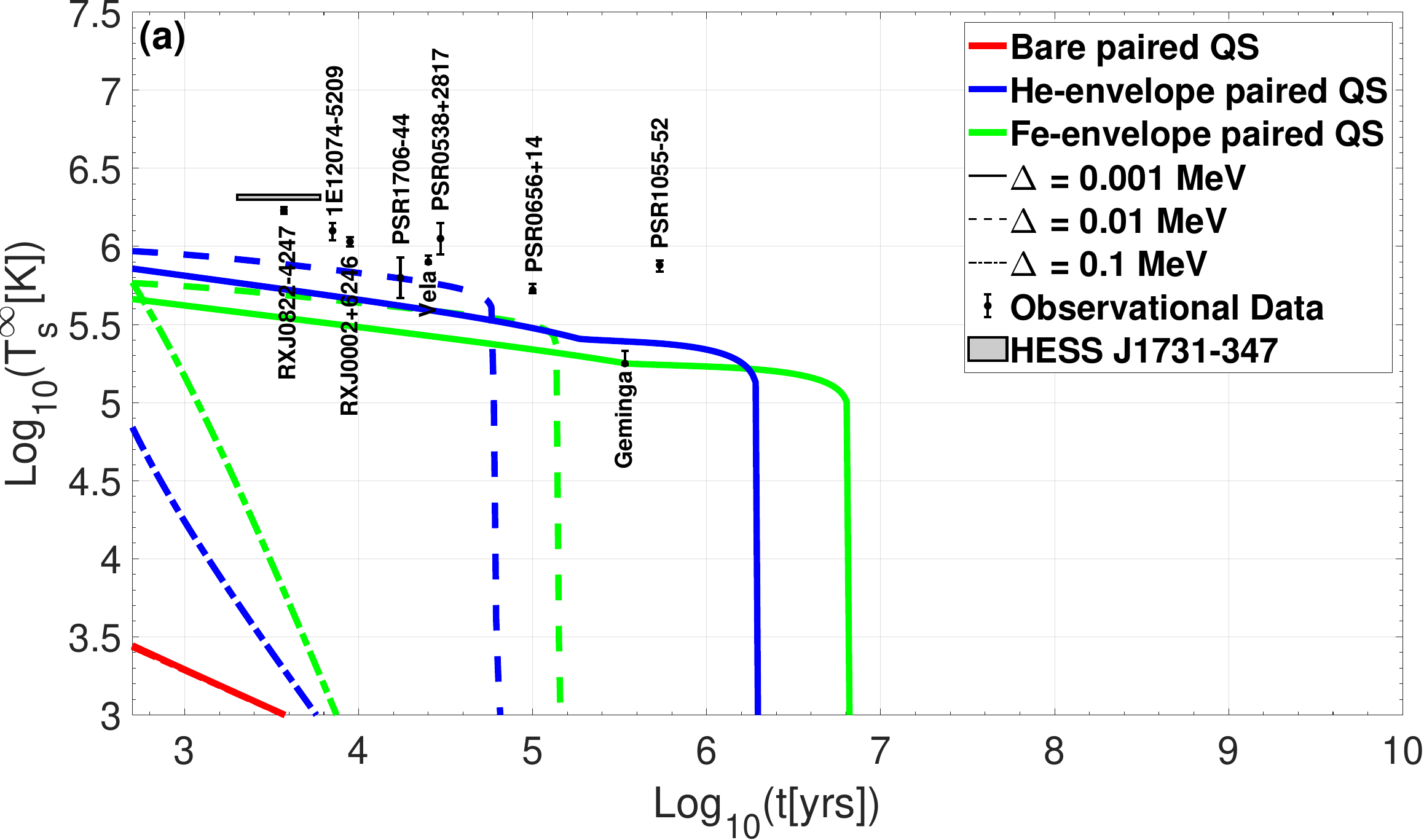}
        \includegraphics[width=0.8\textwidth]{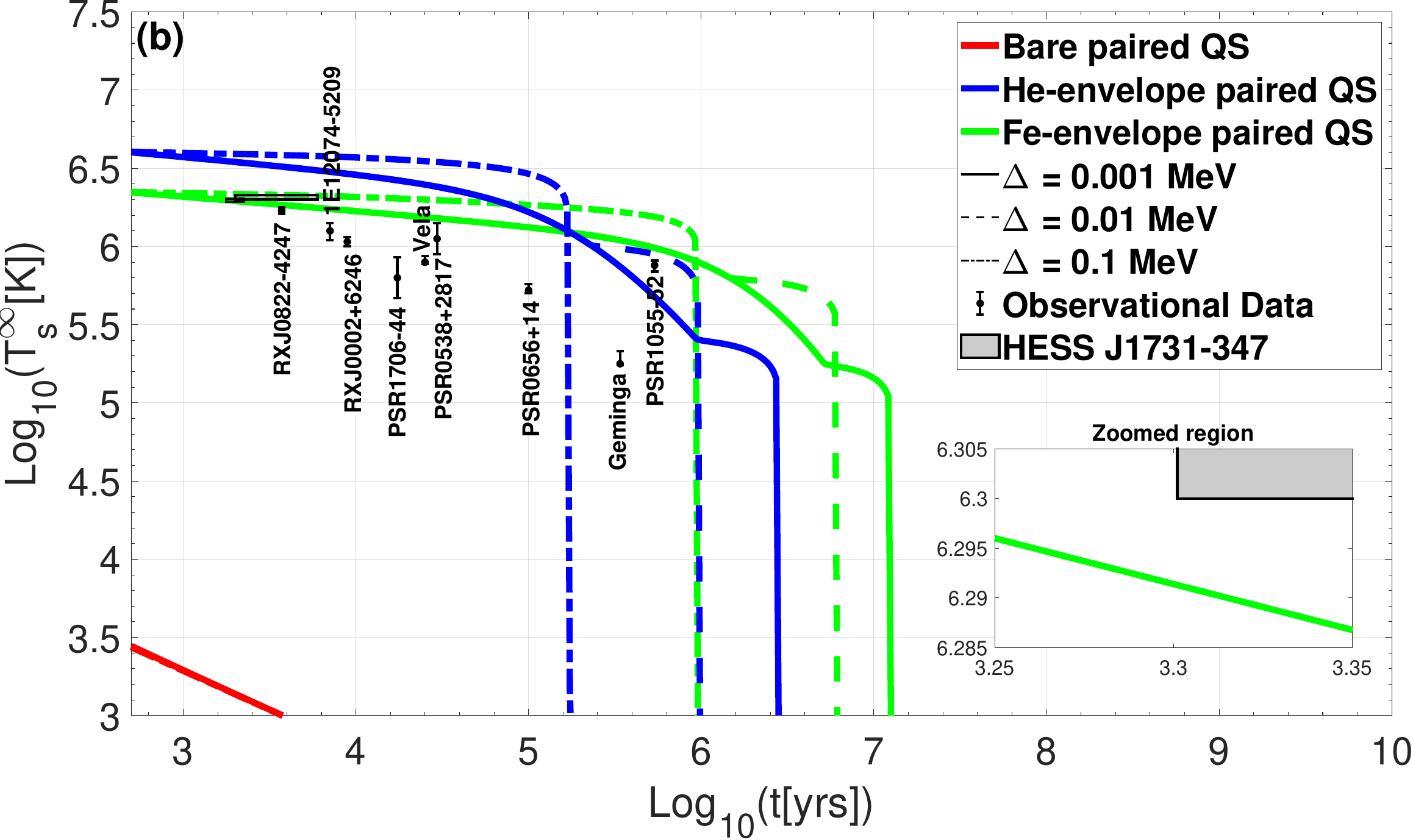}

    \caption{(\textbf{a}) Redshifted surface temperature as a function of time for the QS model, in the case where the quark masses are those of Section~\ref{sec2} (predicting the Fermi momenta of Figure~\ref{fig:fig5}a). Red curves correspond to bare quark stars, blue curves to quark stars with light-element envelopes, and green curves to quark stars with heavy-element envelopes. The solid lines correspond to $\Delta_{0_q} = 0.001$ MeV, the dashed lines to $\Delta_{0_q} = 0.01$ MeV and the dot-dashed lines to $\Delta_{0_q} = 0.1$ MeV.
    (\textbf{b}) Redshifted surface temperature as a function of time for the QS model, considering a combination of quark masses that blocks the d-QDU process (see Appendix~\ref{app:A}). The colors and the styles of the curves denote the same cases of nucleonic envelopes and quark superconductivity as in panel (\textbf{a}).}
    \label{fig:fig6}
\end{figure}

At this point we turn our attention to the case of the QS model. Similarly to the HS scenario, we aimed to study both the CFL and 2SC like pairing patterns. In addition, we included an analysis on the possibility of pairing with low gap values (compared to those expected in an ideal CFL state, see Equation~(\ref{67})).

In Figure~\ref{fig:fig5}a, which depicts the quark Fermi momenta, one can find that the d-QDU process is activated throughout the stellar core. Given that the high temperature of the CCO might be in tension with a thermal evolution driven by the strong d-QDU process, we also focused on studying a scenario where the aforementioned process is blocked. Notably, within the present model, the selection of quark masses plays a crucial role on whether the d-QDU process is activated or not. For instance, we calculated that, for a scenario of equal up and down quark masses (but also other cases as well), d-QDU is blocked. More details on this effect can be found in Appendix~\ref{app:A}. This is particularly interesting because while a slight change in the quark masses has no impact on the QS structure, it might play significant role in its thermal evolution.

Figure~\ref{fig:fig5}b,c contain the results for the thermal evolution of a QS. Apart from considering a model of a bare QS, we also considered the case where a nucleonic heat blanket is present (by utilizing the same relations connecting the surface and internal temperature as before). The red curves in Figures~\ref{fig:fig5}b,c correspond to bare QSs, the blue curves to QSs consisting of an envelope with lighter elements and the green curves to QSs with an envelope of heavier elements. For each of the previous cases, we extracted two different cooling curves each one corresponding to a different superconductivity type (CFL and 2SC). The solid lines present the thermal evolution of the CFL QSs ($\Delta_{0_q} = 15$ MeV) and the dashed lines describe the cooling of QSs when quark matter is in the 2SC phase. In Figure~\ref{fig:fig5}b the d-QDU process is activated, while in Figure~\ref{fig:fig5}c it is blocked due to an alternative selection of quark masses. 

In the case of a CFL star, our calculations for a 15 MeV gap value lead to a thermal evolution with completely suppressed quark heat capacity and reduced neutrino emission. Under the previous consideration, a star cools very rapidly, since it is unable to retain thermal energy due to its low heat capacity. As a consequence, the temperature of the CCO in HESS J1731-347 can not be reconciled for the given age. However, it is important to clarify that the isothermal approximation, employed in this study, has certain caveats in the specific case of CFL QSs with large superconducting gaps. In particular, based on the finding of Ref.~\cite{Blaschke-2000}, CFL stars may exhibit wider thermal relaxation timescales compared to compact stars of different nature (such as NSs, HSs with both low and large gaps, and {paired} QSs with low gaps). When this feature is combined with the fact that the star, due to the suppressed heat capacity, cools rapidly in a timescale comparable to the thermal relaxation timescale, it becomes apparent that results, within the isothermal approximation, may be inaccurate for the early stages of the thermal evolution. In that sense, only solving the full problem would provide a precise description for the earlier stages of a CFL star's thermal evolution, when large pairing gap values are considered. Note, however, that after the end of the thermal relaxation period the star would be expected to cool extremely fast approaching asymptotically the results of the present study (see also Refs.~\cite{Blaschke-2000,Blaschke-2001}).

Moving on to the 2SC QS scenario, the cooling curves presented in Figure~\ref{fig:fig5}b, where the d-QDU is present, fail to reconcile with the HESS J1731-347 constraints regardless of the envelope's composition. In particular, the activation of the strong QDU process dictates rapid cooling scenarios and leads to temperatures lower than the estimated surface temperature of the HESS J1731-347 in the time frame of interest. However, when the d-QDU process is blocked, the 2SC QSs with helium-like and iron-like envelope composition are described by slower cooling scenarios. Specifically, the case of the 2SC QS with Fe-envelope crosses the gray shaded region which corresponds to the age and surface temperature estimation for the CCO (see Figure~\ref{fig:fig5}c). On the contrary, the case of the 2SC QS with He-envelope predicts higher surface temperatures in the specific time frame.

Finally, we performed simulations for lower pairing gap values. As previously mentioned, lower gap values are in tension with the condition of Equation~(\ref{67}) and, therefore, it should be stated that quark matter should not be thought as being in an ideal CFL state. Lower gap values could be helpful to illustrate the suppression related to weaker pairing compared to the fully symmetric CFL superconductivity. Interestingly, as suggested by Refs.~\cite{Horvath-2023, DiClemente-2023} low gap values may be essential for the reconciliation of the HESS J1731-347 constraints.

Similarly to the analysis for large gaps, we considered a scenario where the d-QDU process is activated and one scenario where it is not (see Appendix~\ref{app:A} on the effect of quark masses on the d-QDU process). Figure~\ref{fig:fig6}a corresponds to the case where the d-QDU is activated, while Figure~\ref{fig:fig6}b corresponds to the case where the d-QDU is blocked. The red curves describe bare QSs, the blue curves QSs consisting of an envelope with lighter elements and the green curves QSs with an envelope of heavier elements. The solid lines correspond to $\Delta_{0_q} = 0.001$ MeV, the dashed lines to $\Delta_{0_q} = 0.01$ MeV and the dot-dashed lines to $\Delta_{0_q} = 0.1$ MeV.

The extracted cooling curves appearing in Figure~\ref{fig:fig6}a fail to cross the time--temperature region related to HESS J1731-347. More precisely, the bare QSs cool very rapidly. Then, the consideration of a nucleonic envelope slows down the temperature drop, but it is still not sufficient to counterbalance the effect of the strong d-QDU process. This is not the case in Figure~\ref{fig:fig6}b. The cooling curves describing the cooling evolution of QSs with envelopes constructed by iron-like elements and for $\Delta_{0_q} = 0.001, 0.01$ MeV slightly deviate from the gray shaded area of the HESS J1731-347, while the curve corresponding to $\Delta_{0_q} = 0.1$ MeV appears to be crossing the region of interest. Consequently, the important remark is that the strong neutrino emitting d-QDU process needs to be suppressed in order to achieve agreement between theoretical predictions and observation. Finally, Figure~\ref{fig:fig6}b indicates that the CCO's properties could be explained by unpaired quark matter. This would, however, require the assumption of a nucleonic envelope of heavy elements which conducts heat slightly more efficiently compared to the model employed in this study.

\section{Conclusions}\label{sec5}
In the present study, we considered four different compact star models, each one corresponding to a unique EOS. All models were within the 2$\sigma$ contour related to the mass and radius of the CCO in the HESS J1731-347 remnant. Our main aim was to study their respective thermal evolution to potentially shed light on the possible nature of the aforementioned compact object. 

We initially assumed that the CCO in HESS J1731-347 is a NS, considering two chemical composition scenarios for the envelope and multiple superfluidity models for neutrons and protons. The NS case with an envelope constructed by helium-like elements appeared to be consistent to the surface temperature of the observed compact object for its given age, while the case of the NS with an envelope made up from heavier elements did not cross the temperature-time region of interest. Moreover, the case of a non superfluid NS with envelope consisting of lighter elements was also consistent to the HESS J1731-347 constraints.

Furthermore, we considered the CCO to be a HS. We studied 2 HS models, where in the first case the NDU process was deactivated and in the second one it was activated. We also considered the presence of superfluidity in nuclear matter, implemented two types of superconductivity in quark matter (2SC, CFL) and studied two cases for the envelope's chemical composition. It was found that the reconciliation of the HESS J1731-347 constraints was possible when: (1) the NDU process was not activated, (2) pairing was present in the hadronic phase, (3) the quark matter was in a CFL phase (large gap and pairing between all quarks) and (4) the outermost layers of the star were composed of light (helium-like) elements. Hence, we found that the parametrization of the low-density phase plays a crucial role on the explanation of the CCO's temperature and age. In particular, models stiff enough to enable the NDU process should be ruled out.

Moreover, we examined the possibility that the HESS J1731-347 CCO is either a CFL QS (with $\Delta_{0_q} = 15$ MeV) or a 2SC QS, considering three distinct cases of thermal insulation in the surface of each compact star (no envelope, envelope with light elements, and envelope with heavy elements). We also implemented two different assumptions for the u and d quarks masses which affect the activation of the strong d-QDU process. All cooling simulations for the CFL QS deviated significantly from the region related to the CCO in HESS J1731-347, due to the large pairing gap. As for the 2SC QSs, we could not reach agreement to observational constraints when considering the activation of the d-QDU process. However, when the d-QDU was blocked, the 2SC QS with an external envelope constructed by iron-like elements was consistent with the HESS J1731-347 constraints. The bare 2SC QSs deviated significantly from the region related to the CCO in HESS J1731-347.

Finally, we performed additional cooling simulations considering low gap pairing in quark matter. In the case where the d-QDU process was activated our results were in tension to the observational data on HESS J17311-347. On the contrary, for a gap value of 0.1 MeV, an envelope of heavy elements and no d-QDU activation the temperature and age of the CCO were reconciled.

Future endeavors include studying the cooling profile of the HESS J1731-347 CCO to investigate the impact of other neutrino emitting processes, related to kaon or pion condensation or other exotic degrees of freedom (depending on whether they could appear at the low central densities considered). Moreover, our goal for the upcoming studies is to enhance the present analysis via considering the effects of the magnetic field and the existence of potential heating mechanisms. Furthermore, the present study was restricted to studying abrupt phase transitions. It would be particularly interesting to also examine the impact of considering the presence of a mixed phase (of hadrons and quarks) on our conclusions. In addition, our future target is to re-examine the thermal evolution of the HS, NS, and QS models by numerically solving the general relativistic equations of thermal evolution for a spherically symmetric star~\cite{Thorne-1977}. This will allow us to assess possible effects related to the heat conductivity in dense matter (something potentially important in the QS case with large gap values). 



\vspace{6pt} 

\section*{Author contributions}Conceptualization, D.G.N.,  P.L.-P.,  and C.C.M.; methodology, D.G.N.  and P.L.-P.; software, D.G.N. and P.L.-P.; validation, D.G.N., P.L.-P.,  and C.C.M.; formal analysis, D.G.N. and P.L.-P.; investigation, D.G.N., P.L.-P.,  and C.C.M.; data curation, D.G.N. and P.L.-P.; writing---original draft preparation, D.G.N., P.L.-P.,  and C.C.M.; writing---review $\&$ editing, D.G.N., P.L.-P.,  and C.C.M.; visualization, D.G.N., P.L.-P.,  and C.C.M.; supervision, C.C.M.; project administration, C.C.M.; funding acquisition, P.L.-P. All authors contributed equally to this work. All authors have read and agreed to the published version of the manuscript.

\section*{Funding} 
The research work was supported by the Hellenic Foundation for Research and Innovation (HFRI) under the 5th Call for HFRI PhD Fellowships (Fellowship Number: 19175).

\section*{Data availability} The data associated with this study are
available from the authors upon reasonable request

\section*{Acknowledgments}
We would like to thank the anonymous referees who, by their constructive criticism, significantly improved the present manuscript. P.L.-P. acknowledges that the research work was supported by the Hellenic Foundation for Research and Innovation (HFRI) under the 5th Call for HFRI PhD Fellowships (Fellowship Number: 19175).

\section*{Conflicts of interest}
The authors declare no conflict of interest


\section*{Abbreviations}
The following abbreviations are used in this manuscript:\\

\noindent 
\begin{tabular}{@{}ll}
EOS & Equation Of State \\
SQM & Strange Quark Matter \\
CCO & Central Compact Object \\
QCD & Quantum Chromodynamics \\
QDU & Quark Direct Urca \\
QMU & Quark Modified Urca \\
QB & Quark Bremsstrahlung \\
CFL & Color--Flavor Locked \\
2SC & 2 color--flavor SuperConductor \\
NDU & Nucleon Direct Urca \\
NMU & Nucleon Modified Urca \\
NNB & Nucleon--Nucleon Bremsstrahlung \\
PS & Proton Singlet \\
NT & Neutron Triplet \\
PBF & Pair Breaking Formation \\
QS & Quark Star \\
NS & Neutron Star \\
HS & Hybrid Star \\
HS1 & Hybrid Star-1 \\
HS2 & Hybrid Star-2 \\
env & Envelope \\
\end{tabular}


\appendix

\section{Quark Masses and the Activation of d-QDU} \label{app:A}

The activation of the d-QDU process depends on whether it is kinematically allowed or not. As also mentioned in the main text, the demand for momentum conservation leads to the well-known triangle inequality condition
\begin{equation}\label{a1}
    p_{F_d}<p_{F_u}+p_{F_e},
\end{equation}
which may be written as through Equations~(\ref{21}) and (\ref{22})
\begin{equation}\label{a2}
    n_d^{1/3}<n_u^{1/3} +\bar{n}_e^{1/3},
\end{equation}
with $\bar{n}_e=3n_e$. Notably, since both sides of the inequality are positive, proving that 
\begin{equation}\label{a3}
    n_d^{2/3}<n_u^{2/3}+\bar{n}_e^{2/3} +2(n_u\bar{n}_e)^{1/3},
\end{equation}
would also indicate that the condition of Equation~(\ref{a1}) is satisfied.

One of the equations that can be exploited to make a comparison between all of the involved number densities is the beta equilibrium condition
\begin{equation}\label{a4}
    \mu_d=\mu_u+\mu_e,
\end{equation}
which upon expansion reads 
\begin{equation}\label{a5}
   \sqrt{m_d^2c^4+\mathcal{C} n_d^{2/3}}=\sqrt{m_u^2c^4+\mathcal{C}n_u^{2/3}}+\sqrt{m_e^2c^4+\mathcal{C}\bar{n}_e^{2/3}},
\end{equation}
and $\mathcal{C}=(\hbar\:\pi^{2/3}\: c)^{2}$. Note that terms related to the vector interaction or the density dependence of the bag parameter cancel out. Solving the equation above for $n_d^{2/3}$ gives
\begin{equation}\label{a6}
    n_d^{2/3}=n_u^{2/3}+\bar{n}_e^{2/3}+2n_u^{1/3}\bar{n}_e^{1/3}\sqrt{\frac{m_u^2c^4}{\mathcal{C}n_u^{2/3}}+1}\sqrt{\frac{m_e^2c^4}{\mathcal{C}\bar{n}_e^{2/3}}+1}+\frac{m_u^2+m_e^2-m_d^2}{\mathcal{C}}c^4.
\end{equation}
A close examination of Equation~(\ref{a6}) shows that the difference between the mass of up and down quarks essentially controls the activation of d-QDU. One straightforward example is the scenario where $m_u=m_d$. Then, after performing the relevant cancellation, the right hand side of Equation~(\ref{a6}), will contain the quantity that is supposed to be larger than $n_d^{2/3}$ (from ~(\ref{a3})) and some additional positive terms. Thus, the condition (\ref{a3}) is not satisfied and hence the d-QDU will not be activated. On the contrary, for the selected masses ($m_u=2.3$ MeV and $m_d=4.8$ MeV~\cite{Workmman-2022}) in the main text we found that (\ref{a3}) is guaranteed and, therefore, rapid cooling is allowed.

Since one might argue that the up and down quark masses may not be equal, we need to clarify that we have also numerically checked mass values that are in accordance to the currently accepted ranges~\cite{Workmman-2022} and found pairs that effectively lead to the deactivation of the d-QDU process. It is particularly interesting that while a slight change in the quark masses can have significant impact on the thermal evolution of the star, it does not have any noticeable effect on its structure.

\end{document}